\documentclass[prb,aps,onecolumn,floats,showpacs]{revtex4}

\usepackage{amsmath,amssymb,bm}
\usepackage{graphicx}
\usepackage{color}

\def\eqa{\begin{eqnarray}}
\def\eea{\end{eqnarray}}
\newcommand{\eq}{\begin{equation}}
\newcommand{\ee}{\end{equation}}

\usepackage{physics}

\usepackage[colorlinks=true, linkcolor=blue, citecolor=blue, urlcolor=blue]{hyperref}

\bibpunct{[}{]}{,}{n}{}{}

\begin{document}
\title{Dynamic Kohn anomaly in twisted bilayer graphene}
\author{Jun-Wei Li$^{1}$}
\author{Jia-Xing Zhang$^{1}$}
\author{Wei Chen$^{1,2}$} \email{chenweiphy@nju.edu.cn}
\affiliation{$^{1}$National Laboratory of Solid State Microstructures and School of Physics, Nanjing University, Nanjing, China}
\affiliation{$^2$Collaborative Innovation Center of Advanced Microstructures, Nanjing University, Nanjing, China}

\begin{abstract}
Twisted bilayer graphene (TBG) has attracted great interest in the last decade due to the novel properties it exhibited. It was revealed that e-phonon interaction plays an important role in a variety of phenomena in this system, such as superconductivity and exotic phases. However, due to its complexity, the e-phonon interaction in TBG is not well studied yet. In this work, we study the electron interaction with the acoustic phonon mode in twisted bilayer graphene and one of its consequences, i.e., the Kohn anomaly.
 The Kohn anomaly in ordinary metals usually happens at phonon momentum $q=2k_F$ as a dramatic modification of the phonon frequency when the phonon wave vector nests the electron Fermi surface. However, novel Kohn anomaly can happen in topological semimetals, such as graphene and Weyl semimetals. In this work, we show that the novel dynamic Kohn anomaly can also take place in twisted bilayer graphene due to the nesting of two different Moire Dirac points by the phonon wave vector. Moreover, by tuning the twist angle, the dynamic Kohn anomaly in TBG shows different features. Particularly, at magic angle when the electron bandwidth is almost flat, the dynamic Kohn anomalies of acoustic phonons disappear. We also studied the effects of finite temperature and  doping on the dynamic Kohn anomaly in TBG and discussed the experimental methods to observe the Kohn anomaly in such system.
\end{abstract}

\maketitle  

 \begin{widetext}
\section{Introduction}\label{Sec:I}
Twisted bilayer graphene has intrigued great interest in recent years due to its fascinating  properties, 
such as the unconventional superconductivity \cite{Cao2018,Po2018,XLu2019,Yankowitz_2019,Stepanov2020,Saito2020,Nuckolls2023} at magic angle 
 and the fractional quantum Hall effect at low magnetic 
field\cite{Wilhelm_2021,Xie_2021,Stepanov_2021,Parker_2021,Ledwith_2020,Repellin_2020}. 
Besides, the TBG can also exhibit a variety of exotic 
phases, such as the Chern insulator phase at integer fillings\cite{Cao2018_2,Sharpe2019,Nuckolls2020,Serlin2020,Choi2020,Sharpe2021,Andrei_2021,Jarillo-Herrero_2021,Efetov_2021, Calugaru_2022}
 or the incommensurate Kekule pattern state\cite{Fabrizio_2022, Kwan_2021}.
It was discovered that these exotic phases and properties are often related to the correlation effects originating from the interactions in the system.  
While the role of the e-e Coulomb interaction has been intensively studied in the Chern insulating state of the TBG\cite{Kang_2018,Kang_2019,YHZhang_2019,Senthil_2020,Dai_2021,SongTBG3_2021,SongTBG4_2021,SongTBG5_2021}, the e-phonon interaction  is relatively less studied due to the complexity of such interaction. On the other hand, it was revealed that the e-phonon interaction plays an important role in both the unconventional superconductivity in TBG\cite{Wu_2018,Lian2019,Fabrizio_2019,Guinea_2021,Chen_2023,LiuCX_2023,Dai_2024,Song_Lian_2_2024,Song_Lian_2024}  and some of the exotic state of the  system, such as the Kekule spiral order phase\cite{Kwan_2023}.

A strict and complete study of all the phonon modes in the TBG is unrealistic due to the enormous  number of the phonon modes in this system. Fortunately, only a small percentage of all the phonon modes couples strongly with the electrons and is important\cite{Lian2019,Fabrizio_2019,Guinea_2021,Chen_2023,LiuCX_2023,Dai_2024,Song_Lian_2024,Ochoa_2019,Ishizuka_2021,Wu_2023}. Ref.\cite{Dai_2024} studied the nine dominant optical phonon modes  in TBG and their effects on the electronic states of the system. In this work, we study the e-phonon interaction of the acoustic phonon modes which couple most strongly with the electrons in TBG, which is also considered to be the most important e-phonon coupling for the superconductivity in the TBG\cite{Lian2019,Guinea_2021}.

We derive an effective e-phonon interaction Hamiltonian for the acoustic phonons interacting with electrons in the lowest two bands near the Fermi energy in the continuum model of the TBG. We then studied the dynamic Kohn anomaly~\cite{Kohn1959, Graphene2008} in the system induced by the intervalley e-phonon scatterings. 
The dynamic Kohn anomaly in the continuum model TBG originates from the nesting of the Fermi surface near the Moire Dirac points by the phonon wavevector. Similar phenomena has been observed in 3D Weyl semimetal (WSM) TaP\cite{Nguyen2020} and 2D graphene\cite{Ando2006,
Graphene2008,
Pimenta2009,Jorio2022}. However, we show that TBG exhibits different and more diverse behaviors of the Kohn anomaly due to the adjustable band structure by tuning the twisting angle. 
When the twist angle is away from the magic angle, the Kohn anomaly in TBG exhibits more similarity to that in Weyl semimetals. However, at magic angle when the  two electron bands crossing the Fermi energy become almost flat, the Kohn anomaly induced by the acoustic phonon may disappear because the energy and momentum conservation cannot be satisfied at the same time in the e-phonon scattering processes in the system. The Kohn anomaly in TBG may be observed through inelastic x-ray\cite{Ishikawa2004} and neutron scatterings\cite{Sato1991} in such system.

\section{Review of the Continuum model of the twisted bilayer graphene}

We first have a brief review of  the continuum model of  TBG  following Ref.\cite{Castro2007,MacDonald2011}. We consider a TBG with the two layers  rotated by an angle $\theta/2$ and $-\theta/2$ respectively. Each single layer contains two sublattices labeled by $X=A, B$ in its unit cell. Without lattice distortion, the positions of sublattice $X$ on layer $l$ are given by 
\begin{equation}
\mathbf R^{(l)}_X=m_1 \mathbf a_1^{(l)}+m_2 \mathbf a^{(l)}_2+\boldsymbol \tau^{(l)}_X,
\end{equation}
where $\mathbf a_1^{(l)}$ and  $\mathbf a^{(l)}_2$ are the  lattice unit vectors of graphene layer $l$, 
$m_1, m_2$ are integers and $\boldsymbol \tau^{(l)}_X$ is the sublattice position of atom $X$ in the unit cell in layer $l$. 
For the unrotated graphene, $\boldsymbol \tau^{(l)}_A=(0, 0), \boldsymbol \tau^{(l)}_B=a(0, -\sqrt{3}/3)$ with $a$ the bond length of graphene.

 The Brillouin zones (BZ) of the  single layer graphene and TBG are shown in Fig.1(a). The two Dirac points of the rotated single layer 1 and 2 are denoted as $\mathbf K^{(1)}_\pm$ and $\mathbf K^{(2)}_\pm$ respectively where $+$ and $-$ label the two valleys of a single graphene layer.
 The primitive reciprocal lattice vectors of layer $l$  are ${\mathbf a}^{*(l)}_i=R(\pm \theta/2){\mathbf a}^*_i$ for $l=1,2$ respectively, where ${\mathbf a}_1^*=(2\pi/a)(1,-1/\sqrt{3}), \mathbf{a}^{*}_2=(2\pi/a)(0,2/\sqrt{3})$ are the reciprocal lattice vectors for the unrotated graphene, and $R(\pm \theta/2)$ is the rotation operator. For a given valley, e.g., $\eta=+$, the Moire Reciprocal lattice vectors are given by  $\textbf{g}_{\text{M}i}=\mathbf{a}^{*(1)}_i-\mathbf{a}_i^{*(2)},i=1, 2,$ and the first Moire BZ of the TBG corresponds to the small hexagon in Fig.1a.  We label the two Dirac points of the moire BZ for valley $+$ as  $\mathbf  K_{M1}$ and $\mathbf K_{M2}$.

\begin{figure}
	\includegraphics[width=12
cm]{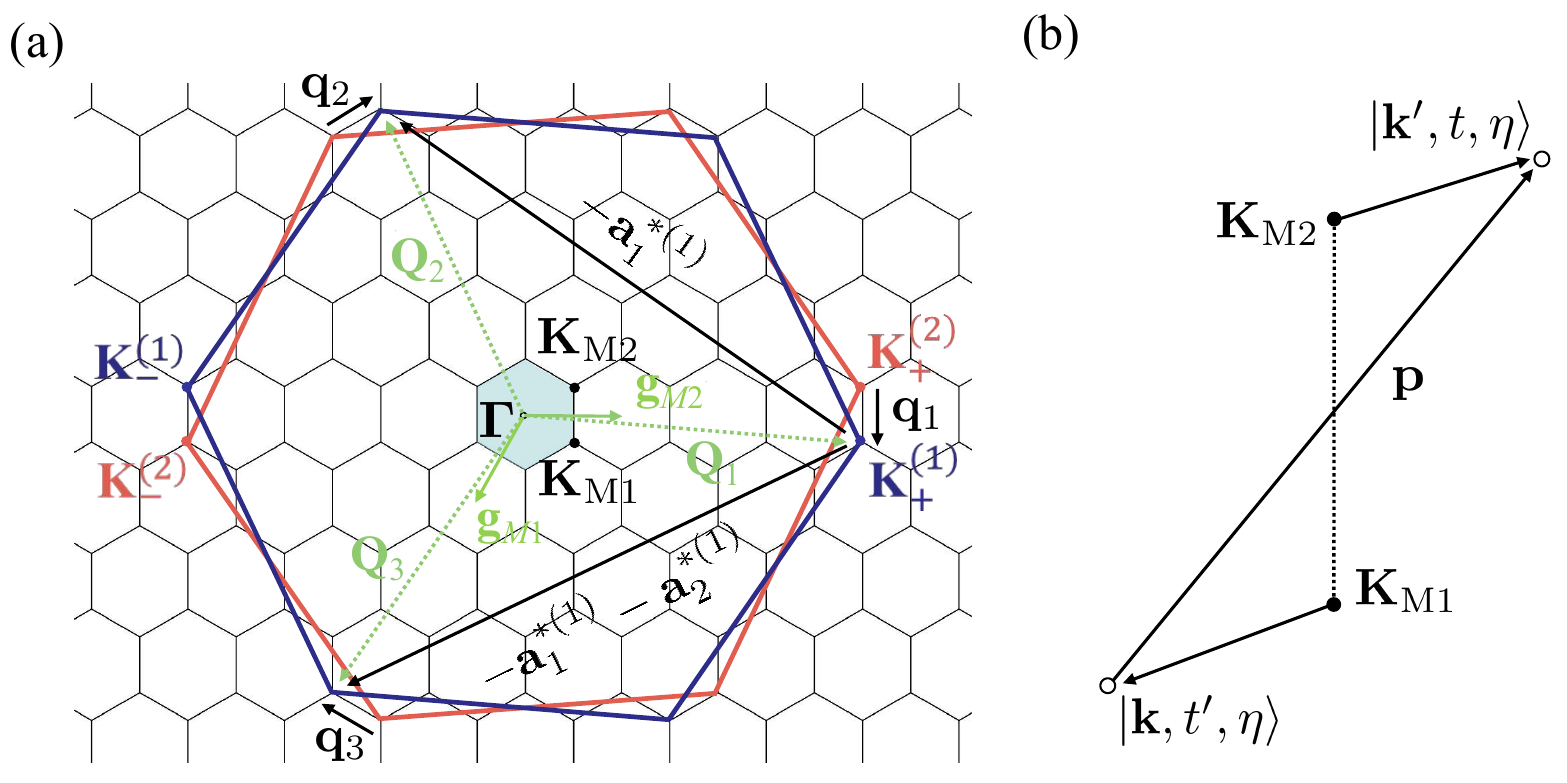}
		\caption{(a)Momentum-space geometry of the  twisted bilayer graphene. The blue and red hexagons mark the first Brillouin zones  of the rotated graphene layer 1 and 2 respectively. $\mathbf K^{(1)}_\pm$ and $\mathbf K^{(2)}_\pm$ label the Dirac points of the two rotated graphene layers and the subscripts $+$ and $-$ label the two valleys of a single graphene layer. The shaded hexagons represent the first Moire BZ of the superlattice. 
		(b)Schematic description of an electron scattering from state $|\mathbf k, t', \eta\rangle$ to $|\mathbf k', t, \eta\rangle$ by a phonon with momentum $\mathbf p$, where $t, t'=\pm$ are the band indices.}\label{fig:NCA_diagram}
\end{figure}

 The electronic band structure of the TBG has been  studied extensively in previous works\cite{Castro2007,MacDonald2011,SongTBG1,Dai_2022,Song_2022,Vishwanath_2019}. We employ the continuum model in Ref.\cite{Castro2007,MacDonald2011} to describe the low energy electronic structure of the TBG in this work. In such model, the hopping processes from the  valley $+$ to $-$ between the two layers are neglected
 due to the large momentum transfer. For  valley $\pm$,  only the three dominant hopping processes near the Dirac points of the two layers 
 with momentum transfer $0$, $\mp{\mathbf g}_{M1}$ and $\mp{\mathbf g}_{M1}\mp{\mathbf g}_{M2}$ 
 between the two layers are retained. The low energy Hamiltonian of the TBG  at valley $\eta$ is then described by the Dirac spinor fields $\psi^{(1)}_{\eta}$ and $\psi^{(2)}_\eta$ of layer $1$ and layer $2$   as: 
 \begin{equation} 
\begin{aligned}
&H_{0, \eta}=\sum_\eta\int d\mathbf{r}
\left(  \psi^{(1)\dagger}_{\eta},\, \psi^{(2)\dagger}_{\eta}  \right)
\begin{pmatrix}
\hbar v_F(-i{\boldsymbol \nabla}-\mathbf{K}^{(1)}_\eta )\cdot  {\boldsymbol \sigma}_{\theta/2}
&  \hat{T}(\mathbf r)  \\
\hat{T}^\dagger(\mathbf r)  &   \hbar v_F(-i{\boldsymbol \nabla}-\mathbf{K}^{(2)}_\eta )\cdot  {\boldsymbol \sigma}_{-\theta/2}
\end{pmatrix}
\begin{pmatrix}  \psi^{(1)}_{\eta}  \\ \psi^{(2)}_{\eta}    \end{pmatrix}.
\end{aligned}
\end{equation}
Here the diagonal matrix elements describe the Hamiltonian of the rotated single  layer 1 and 2 near
the Dirac points and $\boldsymbol\sigma_{\theta/2}=e^{\pm i\theta \sigma_z/4}(\eta \sigma_x, -\sigma_y )e^{\mp i\theta \sigma_z/4}$.
For small rotation angle  $\theta$, $e^{\pm i\theta \sigma_z/4}\approx 1$ and the rotation factor for the single layers can be neglected.

 The off-diagonal matrices $\hat{T}(\mathbf r)$ and $\hat{T}^\dagger(\mathbf r)$ describe the hopping processes between the two layers.  In the continuum model, it is assumed that the interlayer tunneling amplitude between the carbon $p_z$ orbital is a smooth function $T(\mathbf r)$ of the spatial separation projected onto the graphene planes. The tunneling matrix element for an electron with momentum $\mathbf k$ on the sublattice $X$ in  layer 1 to  layer 2 with momentum $\mathbf k'$ on sublattice $X'$ is 
 \begin{equation}
 T_{\mathbf kX}^{\mathbf k'X'}=\bra{\mathbf{k}',X',2}U\ket{\mathbf{k},X,1},
 \end{equation}
where the Bloch states in a single-layer $l$ can be written as 
\eq  
\ket{\mathbf{k},X,l} =
\frac{1}{\sqrt{N}}\sum_{ \mathbf{R}\in\mathbf{R}_X^{(l)}} e^{i\mathbf{k}\cdot\mathbf{R}} 
\ket{ \mathbf{R}+\mathbf{u}^{(l)}_{X}+h^{(l)}_{X}\mathbf{z} }
\ee
with  $\mathbf{u}^{(l)}_{X}$ and $h^{(l)}_{X}\mathbf{z}$ the atomic displacement parallel and perpendicular to the graphene plane respectively.
The interlayer hopping matrix element then becomes
\begin{equation}\label{eq:Hopping_matrix}
\bra{\mathbf{k}',X',2}U\ket{\mathbf{k},X,1}= 
 -\frac{1}{N}\sum_{ \mathbf{R}'\in\mathbf{R}_{X'}^{(2)}}\sum_{ \mathbf{R}\in\mathbf{R}_{X}^{(1)}} e^{i\mathbf{k}\cdot\mathbf{R}-i\mathbf{k}'\cdot\mathbf{R}'} T \left(    \mathbf{R}'+\mathbf{u}^{(2)}_{X'}+h^{(2)}_{X'}\mathbf{z} 
-\mathbf{R}-\mathbf{u}^{(1)}_{X}-h^{(1)}_{X}\mathbf{z} \right),
  \end{equation}
where 
\begin{equation}\label{eq:hopping_integral}
T \left(    \mathbf{R}'+\mathbf{u}^{(2)}_{X'}+h^{(2)}_{X'}\mathbf{z} 
-\mathbf{R}-\mathbf{u}^{(1)}_{X}-h^{(1)}_{X}\mathbf{z} \right)=-\bra{ \mathbf{R}'+\mathbf{u}^{(2)}_{X'}+h^{(2)}_{X'}\mathbf{z} }U \ket{\mathbf{R}+\mathbf{u}^{(1)}_{X}+h^{(1)}_{X}\mathbf{z} }
\end{equation}
only depends on the distance of the two carbon atoms.

At zero atomic displacement, i.e., $\mathbf{u}^{(l)}_{X}=0, \mathbf h^{(l)}_{X}=0$, one can perform the Fourier transform of $T(\mathbf r)$ in Eq.(\ref{eq:Hopping_matrix}) followed by the sum over the lattice sites $\mathbf R$ and $\mathbf R'$ and get 
\begin{equation}\label{eq:T_matrix}
 T_{\mathbf kX}^{\mathbf k'X'}=-\sum_{\mathbf G^{(1)}, \mathbf G^{(2)}}T_{\parallel}(\mathbf k+\mathbf G^{(1)}, d_{X'X})
 e^{-i[\mathbf G^{(1)}\cdot {\boldsymbol \tau}^{(1)}_X-\mathbf G^{(2)}\cdot \boldsymbol \tau^{(2)}_{X'}]} \delta_{\mathbf k+\mathbf G^{(1)}, \mathbf k'+\mathbf G^{(2)}}
\end{equation}
where  $\mathbf G^{(1)}, \mathbf G^{(2)}$ are reciprocal lattice vectors for  layer 1 and layer 2, $d_{X'X}$ is the $z$ direction distance between the nearest $X$ and $X'$ atom in the two layers and
\begin{equation}
T_{\parallel}(\mathbf Q, z)=\frac{ d_0}{2\pi}\int d  p_z T_{\mathbf Q+p_z\hat{\mathbf{z}} } e^{i p_z z}
  = \frac{1}{S_0}\int d^2 r  e^{ -i\mathbf{Q}\cdot \mathbf{r} } T\left( \mathbf{r}+d \hat{\mathbf{z} }\right)
\end{equation}
is the Fourier transform of the hopping matrix in terms of the in-plane coordinates with $T_{\mathbf Q+p_z \hat{\mathbf{z}}}$  the 3D Fourier transform of $T(\mathbf r)$ and $S_0$  the unit cell area.

Since $T_{\parallel}(\mathbf Q, d_{XX'})$ decays quickly with large $Q$, for $\mathbf k$ near the Dirac point $\mathbf K^{(1)}_+$ and $\mathbf k'$ near the Dirac point $\mathbf K^{(2)}_+$,
the dominant coupling occurs in three 
cases $(\mathbf G^{(1)}, \mathbf G^{(2)})=(0, 0), (-\mathbf a^{*(1)}_1, -\mathbf a^{*(2)}_1), (-\mathbf a^{*(1)}_1-\mathbf a^{*(1)}_2, -\mathbf a^{*(2)}_1-\mathbf a^{*(2)}_2)$. The corresponding $\mathbf Q_j\equiv \mathbf k+\mathbf G^{(1)}$ is close to the three equivalent Dirac points of the first BZ of the non-rotated graphene layer $1$. 
In the continuum model, only these  three components of $T_{\parallel}(\mathbf Q_j, d_{XX'})$ are retained and the  matrix elements $T^{\mathbf k'}_{\mathbf k'}$ in Eq.(\ref{eq:T_matrix}) corresponding to these three processes in the sublattice basis are 
\eq   \begin{gathered}   \label{1 hop matr}
T_1=\begin{pmatrix}  t_0 & t_1 \\ t_1 & t_0  \end{pmatrix},  \,
T_2=\begin{pmatrix}  t_0 &  t_1 e^{i \frac{2\pi}{3}} \\ t_1 e^{-i \frac{2\pi}{3}} & t_0  \end{pmatrix},  
T_3=\begin{pmatrix}  t_0 & t_1e^{-i \frac{2\pi}{3}} \\ t_1e^{i \frac{2\pi}{3}} & t_0  \end{pmatrix},
\end{gathered}  \ee
where $t_0=-T_\parallel(\mathbf Q_j, d_{AA}), t_1=-T_\parallel(\mathbf Q_j, d_{AB})$.
Due to the lattice relaxation, the interlayer distance of the AA region becomes greater than the AB region, i.e.,, $d_{AA}>d_{AB}$. For the reason, the hopping energy $t_0$ is smaller than $t_1$. In Ref.\cite{Dai_2022}, the hopping integral $t_1$ is estimated to be $110 \rm meV$ and $t_0\approx 0.8 t_1$.

The electronic Hamiltonian for valley $+$ keeping only the  three $T^i$ matrices  is \cite{Castro2007}
\begin{equation}  \begin{aligned} \label{band H}
&H_{0, \eta=+}=\int d\mathbf{r}
\left(  \psi^{\dag(1)}_{\eta=+},\, \psi^{\dag(2)}_{\eta=+}  \right)
\begin{pmatrix}
\hbar v_F(-i{\boldsymbol \nabla}-\mathbf{K}^{(1)}_+ )\cdot  ( \sigma_x, -\sigma_y ),
&  \sum_{j=1}^3 (T_{j})^\dagger e^{ -i\delta \mathbf{k}_j\cdot\mathbf{r} }
\\
\sum_{j=1}^3 T_{j}
e^{ i\delta \mathbf{k}_j\cdot\mathbf{r} },  &   \hbar v_F(-i{\boldsymbol \nabla}-\mathbf{K}^{(2)}_+ )\cdot  ( \sigma_x, -\sigma_y )
\end{pmatrix}
\begin{pmatrix}  \psi^{(1)}_{\eta=+}  \\ \psi^{(2)}_{\eta=+}    \end{pmatrix},
\end{aligned}  \end{equation}
where $\delta \mathbf{k}_j=0, -\mathbf g_{M_1},  -\mathbf g_{M_1}-\mathbf g_{M_2}$ for $j=1, 2, 3$ respectively and is equal to the momentum transfers between the two layers for the hopping processes.

The Hamiltonian in the momentum space can be truncated using the four single-layer spinor states involved in the three dominant hopping processes, i.e., $|\mathbf k , 1\rangle$ and  $|\mathbf k'=\mathbf k+\delta\mathbf k_i, 2\rangle, i=1, 2, 3$ as basis,  and one gets the truncated Hamiltonian for valley $+$ as \cite{MacDonald2011,SongTBG1}
\begin{equation}\label{eq:Hamiltonian_truncate}
H_{\eta=+,\mathbf{K}_{M1}}^{8\times8}(\mathbf{k}) =
\begin{pmatrix}
v_F(\mathbf{k}-\mathbf K^{(1)}_+)\cdot\boldsymbol{\sigma}^{*} & T_1^\dagger & T_2^\dagger & T_3^\dagger 
\\
T_1 & v_F(\mathbf{k}+\delta\mathbf k_1-\mathbf K^{(2)}_+)\cdot\boldsymbol{\sigma}^{*} & 0 & 0 
\\
T_2 & 0 & v_F(\mathbf{k}+\delta\mathbf k_2-\mathbf K^{(2)}_+)\cdot\boldsymbol{\sigma}^{*} & 0 
\\
T_3 & 0 & 0 & v_F(\mathbf{k}+\delta\mathbf k_3-\mathbf K^{(2)}_+)\cdot\boldsymbol{\sigma}^{*} 
\end{pmatrix},
\end{equation}
where $\boldsymbol\sigma^*=(\sigma_x, -\sigma_y)$ acts on the sublattice basis.

The Moire momentum $\tilde{\mathbf k}$ characterizing the band structure of the TBG  can be obtained by zone folding of the single layer momentum $\mathbf k$ or $\mathbf k'$ to the Moire BZ, i.e., $\mathbf k=\tilde{\mathbf k}+m_1\mathbf g_{M1}+m_2 \mathbf g_{M2}, \mathbf k'=\tilde{\mathbf k}+n_1\mathbf g_{M1}+n_2 \mathbf g_{M2}$, where $m_1, m_2, n_1, n_2$ are integers and $\tilde{\mathbf k}$ locates in the Moire BZ. 
The Hamiltonian Eq.(\ref{eq:Hamiltonian_truncate}) gives eight effective low energy bands for the TBG. There are two bands per spin and per valley in the middle of the spectrum that touch at two inequivalent  Moire Dirac points $\mathbf K_{M1}$ and $\mathbf K_{M2}$ 
shown in Fig.1(a) with a linear dispersion due to the $\hat{\mathcal{C}}_{2z}\hat{\mathcal{T}}$ and $\hat{\mathcal{C}}_{3z}$ symmetry\cite{Wu_2018}. At neutrality, the Fermi energy locates at the Moire Dirac points. An effective $2*2$ Hamiltonian for the  two bands near the Moire Dirac point  $\mathbf K_{M1}$ was constructed by the $\mathbf k\cdot \mathbf p$ expansion near $\mathbf K_{M1}$ as\cite{MacDonald2011,SongTBG1} 
\begin{equation}\label{eq:effective_H}
H^*(\tilde{\mathbf k})= \hbar v (\tilde{\mathbf{k}}-\mathbf K_{M1})\cdot \boldsymbol{\sigma}^{*}
\end{equation}
where $v = \frac{1-3\alpha_1^2}{1+3\alpha_0^2+3\alpha_1^2}v_F$ with $\alpha_0 = \frac{t_0}{\hbar v_F \abs{\mathbf{q}_1} }, \alpha_1 = \frac{t_1}{\hbar v_F \abs{\mathbf{q}_1} }$. The effective Hamiltonian near ${\mathbf K}_{M2}$ has the same form by replacing ${\mathbf K}_{M1}$ by ${\mathbf K}_{M2}$. The two Moire Dirac points $\mathbf K_{M1}$ and $\mathbf K_{M2}$ carry the same chirality\cite{Castro2011,He2013,Po2019} because they originate from the unperturbed Dirac cones of the two different layers in the same valley.

\section{Electron-phonon interaction Hamiltonian in the TBG}
TBG contains a large number of phonon modes due to the large number of atoms in a unit cell. However, only a limited number of phonon modes couple strongly with the electrons. Ref.\cite{Dai_2024} studied the nine dominant optical phonon modes which couple strongly with the electrons and their effects on stablizing the electron order in the system. In this work, we instead focus on the most important acoustic mode, i.e.,  the in-plane $\mathbf u^-$ mode in TBG  in the following text, which was considered to be the dominant phonon mode resulting in the superconductivity in TBG\cite{Lian2019}.

\subsection{Interlayer e-phonon scattering matrix}

The interlayer electron-phonon interaction Hamiltonian in TBG can be
obtained by expanding the hopping matrix elements  in Eq.(\ref{eq:Hopping_matrix}) in terms of the  lattice displacement $\mathbf u_X^{(l)}$ and $h_X^{(l)}$, as shown in Ref.\cite{Koshino2020,LiuJP2023}. Since the flexural mode $h_X^{(l)}$  perpendicular to the graphene plane has quadratic dispersion\cite{Mariani2008} and is weak at low energy. We only consider the in-plane mode $\mathbf u_X^{(l)}$ in the following. 
We absorb the static lattice relaxation effects to the hopping integral $t_0$ and $t_1$ and only consider the expansion in terms of the dynamic vibration of the lattice. It was shown that the static lattice relaxation does not change the electronic band structure qualitatively~\cite{Koshino2020}. 

  Since the interlayer hopping energy only depends on the relative displacement of the two layers, i.e., $\mathbf u^-=\mathbf u^{(2)}-\mathbf u^{(1)}$, we only consider the dominant acoustic phonon modes corresponding to $\mathbf u^-$  and neglect the centre of mass modes $\mathbf u^+=(\mathbf u^{(2)}+\mathbf u^{(1)})/2$.
  
At finite displacement $\mathbf u^{(l)}_X$, the interlayer hopping matrix element Eq.(\ref{eq:Hopping_matrix}) after the Fourier transform of $T({\mathbf r})$ becomes
\begin{equation}
\bra{\mathbf{k}',X',2}U\ket{\mathbf{k},X,1}=-\frac{1}{N}\frac{S_0 d_0}{(2\pi)^3}\int d^3 \tilde{\mathbf p} T(\tilde{\mathbf p})\sum_{\mathbf R\in  \mathbf R^{(1)}_X, \mathbf R'\in \mathbf R^{(2)}_{X'}}e^{i(\mathbf k-\tilde{\mathbf p})\mathbf R-i \tilde{\mathbf p}\cdot \mathbf u^{(1)}_X(\mathbf R)}
e^{-i(\mathbf k'-\tilde{\mathbf p})\mathbf R'+i \tilde{\mathbf p}\cdot \mathbf u^{(2)}_{X'}(\mathbf R')}.
\end{equation}

Following the procedure in Ref.\cite{Koshino2020}, i.e.,  replacing $u^{(l)}_X$ by its Fourier transform
\eq  \label{lat dist F trans}
\mathbf{u}_{X}^{(l)}(\mathbf{r})=\frac{1}{\sqrt{S}}\sum_{\mathbf{p}} \mathbf{u}_{X,\mathbf{p}}^{(l)}e^{i\mathbf{p}\cdot \mathbf{r}},\quad   
\ee
and expanding the exponential functions $\exp(i\tilde{\mathbf p}\cdot \mathbf u_{\mathbf p} e^{i\mathbf p\cdot \mathbf R})$ in a Taylor series, one gets the interlayer hopping matrix element after the sum over the lattice sites $\mathbf R$ and $\mathbf R'$ as
\eq  \begin{aligned}   \label{eq:Hopping_matrix_2}
&    \bra{\mathbf{k}',X',2}U\ket{\mathbf{k},X,1}
\\  =  & \sum_{\mathbf{G}^{(1)},\mathbf{G}^{(2)}}\sum_{n_1,n_2,..}\sum_{n'_1,n'_2,..}\Gamma^{(n'_1,n'_2..)}_{(n_1,n_2,..)}(\mathbf{Q})
e^{-i(\mathbf{G}^{(1)}\cdot\boldsymbol{\tau}^{(1)}_{X}-\mathbf{G}^{(2)}\cdot\boldsymbol{\tau}^{(2)}_{X'})}
\delta_{\mathbf{k}+\mathbf{G}^{(1)}+n_1\mathbf{p}_1
..,\ \mathbf{k}'+\mathbf{G}^{(2)}-n'_1\mathbf{p}_1-..}
\end{aligned}  \ee 
where $\mathbf p_{1,2}$  are the momentum  from the Fourier transform of $\mathbf u^{(l)}_X$,
and $\mathbf{Q}=\mathbf{k}+\mathbf{G}^{(1)}+n_1\mathbf{p}_1+n_2\mathbf{p}_2+..=\mathbf{k}'+\mathbf{G}^{(2)}-n'_1\mathbf{p}_1- n'_2\mathbf{p}_2-..$, with  $n_i, n'_i=0,1,2...$. The Fourier component of the hopping matrix
\begin{equation}  \begin{aligned}  
\Gamma^{(n'_1, n'_2,..)}_{(n_1, n_2,..)}(\mathbf{Q})=   &-
\frac{ d_0}{2\pi} \int d  p_z \, T_{\mathbf{Q}+p_z\hat{\mathbf{z}} } e^{i p_z d_{X'X}}  \frac{\left[ -i\mathbf{Q}\cdot\mathbf{u}^{(1)}_{X,\mathbf{p}_1}\right]^{n_1}}{n_1!(\sqrt{S})^{n_{1}}}\frac{\left[ -i\mathbf{Q} \cdot \mathbf{u}^{(1)}_{X,\mathbf{p}_2}\right]^{n_2}}{n_2!(\sqrt{S})^{n_{2}}}...
\frac{\left[ i\mathbf{Q}\cdot\mathbf{u}^{(2)}_{X',\mathbf{p}_1}\right]^{n'_1}}{n'_1!(\sqrt{S})^{n'_{1}}}\frac{\left[ i\mathbf{Q}\cdot\mathbf{u}^{(2)}_{X',\mathbf{p}_2}\right]^{n'_2}}{n'_2!(\sqrt{S})^{n'_{2}}}...
\end{aligned}  
\end{equation}

For small twist angles  and long range displacement, $u^{(l)}_A(\mathbf r)\approx u^{(l)}_B(\mathbf r)=u^{(l)}(\mathbf r)$, where $u^{(l)}(\mathbf r)$ is smooth varying compared to the atomic scale, $\mathbf p$ in $\mathbf u^{(l)}_{\mathbf p}$ is much smaller than $1/a$ and one can neglect the small shift $n_1\mathbf p_1+n_2 \mathbf p_2+...$ in $\mathbf Q$ in the coupling amplitude $\Gamma^{(n'_1, n'_2,..)}_{(n_1, n_2,..)}(\mathbf{Q})$ and replace $\mathbf Q$ by the three ${\mathbf Q}_j$ in the above text, i.e.,
$\mathbf Q_1= \mathbf K_+,\  \mathbf Q_2=\mathbf K_+-\mathbf a^*_1,\  \mathbf Q_3=\mathbf K_+-\mathbf a^*_1-\mathbf a^*_2.$ 
In the leading order we only need to keep the linear order of 
 the displacement $\mathbf u_{\mathbf p}$ and get the e-phonon interaction matrix element as
\begin{eqnarray}\label{eq:e-phonon}
 && \bra{\mathbf{k}',X',2}H_{e-ph}\ket{\mathbf{k},X,1} \nonumber\\ 
    & =&   -\frac{1}{\sqrt{S}}\sum_{\mathbf{p}} \sum_{j=1,2,3} T_\parallel\left( \mathbf{Q}_j;\,d_{X'X}\right) e^{i(-\mathbf{G}_j^{(1)}\cdot\boldsymbol{\tau}^{(1)}_{X}+\mathbf{G}_j^{(2)}\cdot\boldsymbol{\tau}^{(2)}_{X'})}\left(i\mathbf{Q}_j\cdot \mathbf{u}^{-}_{\mathbf{p}}\right)\delta_{\mathbf{k}+\delta \mathbf{k}_j+\mathbf{p},\mathbf{k}'}, 
\end{eqnarray}
where $\mathbf u^-_{\mathbf p}$ is  the Fourier transform of $\mathbf u^-(\mathbf r)$.
This matrix element can be written in the sublattice basis as
\eq \begin{aligned}\label{eq:e-ph_matrix}
  \bra{\mathbf{k}',2}H_{e-ph}\ket{\mathbf{k},1} = 
 \frac{1}{\sqrt{S}}\sum_{\mathbf{p}} \sum_{j=1,2,3} T_j\left(i\mathbf{Q}_j\cdot\mathbf{u}^{-}_{\mathbf{p}}\right)\delta_{\mathbf{k}+\delta \mathbf{k}_j+\mathbf{p},\mathbf{k}'},    
\end{aligned}  \ee
where $T_i, i=1,2, 3$ is defined in the last section.

\subsection{Intra-layer e-phonon scattering matrix} 

For e-phonon scatterings within the same single-layer, we neglect the weak interaction between the two layers, and the electron-phonon Hamiltonian is identical to that in single-layer graphene. 
In the continuum limit, the matrix element is given by\cite{Ando2002,Pereira2009,Guinea2010} 
\begin{equation}   \label{intra_ep_matrix}
 \bra{ \mathbf{k}',l  } H_{e-ph}
    \ket{ \mathbf{k},l }  =
 e v_F \left( \sum_{\mathbf{p}}\mathbf{A}^{(l)}_{\eta,\mathbf{p}} \delta_{ \mathbf{k}',\mathbf{k}+\mathbf{p} } \right) \cdot (\eta \sigma_x, -\sigma_y ) ,
\end{equation}
where $\mathbf{A}^{(l)}_{\eta}$ represents the pseudo-vector potential resulting from in-plane lattice displacement. 
The out-of-plane single layer phonon mode couples quadratically to the electrons\cite{Mariani2010,Ochoa2012} and does not contribute to the intra-layer e-phonon scattering in the first-order so we neglect it.
The single layer pseudo-vector potential is expressed as\cite{Ando2002,Pereira2009,Guinea2010} 
\begin{equation}
 \mathbf{A}^{(l)}_{\eta,\mathbf{p}}  =  
- \frac{3}{4} \eta \frac{\beta_0 \gamma_0}{e v_F}   ( \hat{W}_{\mathbf{p}} \, \mathbf{u}^{(l)}_{\mathbf{p}} ) .
\end{equation}
Furthermore, $\beta_0 \approx 3.14$  remains constant\cite{Koshino2020}, and $\gamma_0=2.8\,\text{eV}$ represents the electronic nearest hopping energy in single-layer graphene. 
The matrix $\hat{W}_\mathbf{p}$ is given by\cite{Ando2002,Pereira2009,Guinea2010}
\begin{equation}
\hat{W}_\mathbf{p}=\matrixquantity( i p_x &  -i p_y
\\ -i p_y  &   -i p_x )  
\end{equation}

The in-plane phonon modes can be divided into the longitudinal mode with polarization direction $\hat{\mathbf e}_{\mathbf p, L}=\mathbf p/|\mathbf p|$ and transverse mode with polarization direction $\hat{\mathbf e}_{\mathbf p, T}=\hat{\mathbf e}_z\cross \mathbf p/|\mathbf p|$. The atomic displacement can then be quantized as\cite{Lian2019}
\begin{eqnarray} 
\mathbf{u}^{-}_{\mathbf{p}} &=&\sum_{\chi=T,L}
 \sqrt{ \frac{\hbar}{\rho\omega_{0,\mathbf{p},\chi}} }i\hat{\mathbf{e}}_{\mathbf{p},\chi}
 \left(  \hat{a}_{\mathbf{p},\chi}+\hat{a}^{\dagger}_{-\mathbf{p},\chi}    \right), 
\end{eqnarray}
where  $\hat{a}_{\mathbf{p},\chi}$ is the in-plane phonon annihilation operator with momentum $\mathbf p$ and polarization $\chi=L, T$.
The phonon frequency $\omega_{0, \mathbf p, \chi}=c_\chi |\mathbf p|$ for $\chi=L, T$. For simplicity\cite{Ando2002,Lian2019}, we  take $c_L=c_T=s=10^{4}\text{ m/s}$.

\subsection{Projection of e-phonon interaction to the lowest two bands} 

 At low energy, we only need to consider the two bands near $E = 0$. The effective Hamiltonian for this two-band system is given in Eq.(\ref{eq:effective_H}) in the  sub-lattice basis. The two eigenstates of the two-band system near the Moire Dirac point ${\mathbf K}_{M1}$ can be solved as
\begin{eqnarray}
 \ket{ \bm{\phi}^{s=+}_{\eta=+}(\tilde{ \mathbf{k} })} &=&
\frac{1}{\sqrt{1+3\alpha_0^2+3\alpha_1^2}} \left[
\ket{ \mathbf{k},A,1 }
  - \sum_{j=1,2,3}\left(
-i\alpha_1\ket{ \mathbf{k}+\delta \mathbf k_j,A,2 } 
+i\alpha_0e^{-i\frac{2\pi}{3}(j-1)}\ket{ \mathbf{k}+\delta \mathbf k_j,B,2 } 
\right) \right], \label{Dirac_basis_1}
 \\
 \ket{ \bm{\phi}^{s=-}_{\eta=+}(\tilde{ \mathbf{k} })} &=&
\frac{1}{\sqrt{1+3\alpha_0^2+3\alpha_1^2}} \left[
\ket{ \mathbf{k},B,1 }
  - \sum_{j=1,2,3}\left(
-i\alpha_0e^{i\frac{2\pi}{3}(j-1)}\ket{ \mathbf{k}+\delta \mathbf k_j,A,2 } 
+i\alpha_1\ket{ \mathbf{k}+\delta \mathbf k_j,B,2 } 
\right) \right].
 \end{eqnarray}
Similarly, the eigenstates near the Moire Dirac point ${\mathbf K}_{M2}$ are solved as
 \begin{eqnarray}  
 \bra{ \bm{\phi}^{s'=+}_{\eta=+}(\tilde{ \mathbf{k} }')} &=&
\frac{1}{\sqrt{1+3\alpha_0^2+3\alpha_1^2}} \left[
\bra{ \mathbf{k}',A,2 }
  + \sum_{m=1,2,3}\left(
i\alpha_1\bra{ \mathbf{k}'-\delta \mathbf k_m,A,1 } 
-i\alpha_0e^{i\frac{2\pi}{3}(m-1)}\bra{ \mathbf{k}'-\delta \mathbf k_m,B,1 } 
\right) \right],
 \\
 \bra{ \bm{\phi}^{s'=-}_{\eta=+}(\tilde{ \mathbf{k} }')} &=&
\frac{1}{\sqrt{1+3\alpha_0^2+3\alpha_1^2}} \left[
\bra{ \mathbf{k}',B,2 }
  +\sum_{m=1,2,3}\left(
i\alpha_0e^{-i\frac{2\pi}{3}(m-1)}\bra{ \mathbf{k}'-\delta \mathbf k_m,A,1 } 
-i\alpha_1\bra{ \mathbf{k}'-\delta \mathbf k_m,B,1 } 
\right) \right] \label{Dirac_basis_2}
 \end{eqnarray}
where
$\mathbf k, \mathbf k'$ are the single layer momentum near $\mathbf K^{(1)}_+$ and $\mathbf K^{(2)}_+$ respectively, and $\tilde{\mathbf k}, \tilde{\mathbf k'}$ are the Moire momentum band-folded from the single layer momentum $\mathbf k$ and $\mathbf k'$.
At $\mathbf k=\mathbf K^{(1)}_+$, i.e., $\tilde{\mathbf k} = \mathbf K_{M1}$,  $|\phi^\pm_+ (\tilde{ \mathbf{k}  })\rangle $   
give
the two degenerate eigenstates of Eq.(\ref{eq:Hamiltonian_truncate}) with $E = 0$. Similarly, at $\mathbf k'=\mathbf K^{(2)}_+$, i.e.,  $\tilde{\mathbf k}'=\mathbf K_{M2}$,  $|\phi^\pm_+ (\tilde{ \mathbf{k}'  })\rangle$ give the two degenerate eigenstates with $E=0$ at $\mathbf K_{M2}$.

At zero doping, the Fermi surface aligns  with the Moire Dirac points $\mathbf{K}_{\text{M1}}$ and $\mathbf{K}_{\text{M2}}$ of the two bands crossing $E=0$.
We then focus on the inter-Moire-Dirac-point scatterings induced by phonons, which may result in a dynamic Kohn anomaly due to the nesting of the Fermi surface.  Considering the e-phonon scatterings in the valley $+$, the scattering matrix from the two eigenstates $|\phi^\pm_+(\tilde{\mathbf k})\rangle$ near $\mathbf K_{M1}$ to the two eigenstates $|\phi^\pm_+(\tilde{\mathbf k}')\rangle$ near $\mathbf K_{M2}$ can be obtained from Eq.(\ref{eq:e-ph_matrix}), (\ref{intra_ep_matrix}), (\ref{Dirac_basis_1})-(\ref{Dirac_basis_2}) and written in the sublattice basis 
 as
\begin{equation}  \label{eq:int-D-ep}
    \bra{ \bm{\phi}^{s'}_{+}( \tilde{\mathbf{k}}' ) } H_{e-ph} \ket{ \bm{\phi}^s_{+}( \tilde{\mathbf{k}} ) }= \frac{1}{\sqrt{S}} 
   \sum_{\chi}\sum_{\mathbf{p}} [\hat{g}_{+,\chi}(\mathbf{p} )]_{s's}\delta_{\tilde{\mathbf k}', \tilde{\mathbf k}+\mathbf p+m\mathbf g_{\mathbf M1}+n \mathbf g_{\mathbf M2}}
(a_{\mathbf{p} ,\chi}+a^{\dagger}_{-\mathbf{p},\chi}),
\end{equation}
where $\tilde{\mathbf k'}$ and $\tilde{\mathbf k}$ are the Moire momentum near $\mathbf K_{M1}$ and $\mathbf K_{M2}$ respectively, $\mathbf p$ is the phonon momentum and the $\delta$ function gives the momentum conservation during the e-phonon scatterings. 
The Umklapp processes are subleading due to the large mismatch of energy and momentum so we only consider the case $m=0, n=0$.

 For $\tilde{\mathbf k} (\tilde{\mathbf k}')$ locating near $\mathbf K_{M1} (\mathbf K_{M2})$ or its two other equivalent Moire Dirac points,
the phonon momenta for the inter-Moire-Dirac-point scattering then take the values $\mathbf p\approx -\mathbf q_m, m=1, 2, 3$. The scattering matrices with these three phonon momenta are connected by the ${\cal C}_{3z}$ symmetry. The scattering matrix $\hat{g}_{+,\chi}(\mathbf p)$ includes the contribution from both inter-layer and intra-layer scatterings, i.e., $\hat{g}_{+,\chi}(\mathbf p) =\hat{g}^{\rm inter}
_{+, \chi}(\mathbf p) +\hat{g}^{\rm intra}_{+, \chi} (\mathbf p)$.  For $\mathbf p\approx -\mathbf q_1$, 
 we get the interlayer scattering matrix in the sublattice basis as
\begin{eqnarray}  
 && \hat{g}^{\rm inter}_{\eta=+,\chi=T,L}(\mathbf p)=
\nonumber\\
&&
- \xi \sqrt{\frac{\hbar}{\rho\omega_{0,\mathbf{p},\chi}}} \left[
\hat{T}_1 (\mathbf{Q}_1 \cdot \hat{\mathbf{e}}_{\mathbf{p},\chi} )  
-    \sum_{i,j,n=1, 2, 3}  
 \left( \frac{\hat{T}_{i}^\dagger}{\hbar v_F\abs{\mathbf{q}_1} } \right) \left( \frac{ \mathbf{q}_{i}\cdot\boldsymbol{\sigma}^{*\dagger} }{\abs{\mathbf{q}_1}} \right) 
 \hat{T}_n\left( \frac{-\mathbf{q}_j\cdot\boldsymbol{\sigma}^{*}}{\abs{\mathbf{q}_1}}  \right)\left( \frac{\hat{T}_j}{\hbar v_F\abs{\mathbf{q}_1} } \right) ( \mathbf{Q}_n \cdot \hat{\mathbf{e}}_{\mathbf{p},\chi} )
 \right],
 \end{eqnarray}
where $\chi= T, L$ represent the transverse and longitudinal phonon modes respectively, $\boldsymbol{\sigma}^{*}$ acts on the sublattice basis,  $\xi=\frac{ 1}{1+3\alpha_0^2+3\alpha_1^2}$ and ${\hat{\mathbf e}}_{\mathbf{p},\chi}$ 
is the polarization direction of the phonon modes. 
For $\mathbf p\approx-\mathbf q_1$, the momentum conservation restricts $(i, j, n)$ only to  the values $(1,1,1), (1,2,2)$, $(2,1,2)$, $(1,3,3)$, $(3,1,3)$.

The contribution to the e-phonon matrix elements Eq.(\ref{eq:int-D-ep}) from the intralayer e-phonon scatterings  is obtained as
\begin{eqnarray}
  && \hat{g}^{\rm intra}_{\eta=+,\chi=T,L}(\mathbf{p}) =
  \nonumber\\
 &&  \frac{3}{8}\xi\beta_0 \gamma_0 \sqrt{ \frac{ \hbar }{\rho \omega_{0,\mathbf{p},\chi} } }\bigg[     
  \left( \frac{\hat{T}_{1}^\dagger}{\hbar v_F\abs{\mathbf{q}_1} } \right) 
\left( \frac{ \mathbf{q}_{1}\cdot\boldsymbol{\sigma}^{*\dagger} }{\abs{\mathbf{q}_1}} \right) 
\Big(  ( i \hat{W}_{\mathbf{p}} \, \hat{\mathbf{e}}_{\mathbf{p},\chi} )     \cdot \boldsymbol{\sigma}^{*}      \Big)   
 -       \Big(  ( i \hat{W}_{\mathbf{p}} \, \hat{\mathbf{e}}_{\mathbf{p},\chi} )     \cdot \boldsymbol{\sigma}^{*}      \Big)\left( \frac{-\mathbf{q}_1\cdot\boldsymbol{\sigma}^{*}}{\abs{\mathbf{q}_1}}  \right)\left( \frac{\hat{T}_1}{\hbar v_F\abs{\mathbf{q}_1} } \right)
 \bigg]. \nonumber\\
\end{eqnarray}

Putting the contributions from the interlayer and intralayer e-phonon scatterings together, we get the  effective e-phonon interaction Hamiltonian for the inter-Moire-Dirac-Point scatterings in the effective two band system as 
\begin{equation}  \begin{aligned}  \label{eq:two_band_ep}
H^{*}_{\rm e-ph}  =  &\frac{1}{\sqrt{S}} 
\sum_{\tilde{\mathbf k},\eta}\sum_{\mathbf{p}\approx-\mathbf{q}_m,\chi}
\bm{\hat{\phi}}^{\dagger}_{\eta}(  \tilde{\mathbf k}+\mathbf{p} ) 
    \cdot \left[ 
\hat{g}_{\eta,\chi}(\mathbf{p})\left( 
a_{\mathbf{p},\chi}+a^{\dagger}_{-\mathbf{p},\chi}   \right) 
 \right] \cdot \bm{\hat{\phi}}_{\eta}( \tilde{\mathbf k} )  
 + \text{h.c.},
 \end{aligned}  \end{equation} 
where $\bm{\hat{\phi}}$ is the electron field of the two band system,  and $\hat{g}_{\eta,\chi}(\mathbf{p})$ for $\eta=+, \mathbf p\approx -\mathbf q_1$ can be written as 
\begin{eqnarray}
 \hat{g}_{\eta=+,T}(\mathbf{p}\approx-\mathbf{q}_1)  &=&  
 g_{T,\mathbf{p}} 
 \left(  
  \alpha_0 \sigma_0  +   \alpha_1 \sigma_x 
\right), \\
  \hat{g}_{\eta=+,L}(\mathbf{p}\approx-\mathbf{q}_1)  &=&  
  g_{L,\mathbf{p}}  
 \alpha_1 \sigma_y,
\end{eqnarray}
where $\sigma_0, \sigma_x, \sigma_y$ are the Pauli matrices acting on the sublattice basis and  the
coupling constant 
\begin{equation}
g_{T,\mathbf{p}} =    \frac{ \tilde{g}^{\rm intra}_{T,\mathbf{p}} + \tilde{g}^{\rm inter}_{T,\mathbf{p}}(1 + 2\alpha_0^2  -2\alpha_1^2)  }{1+3\alpha_0^2+3\alpha_1^2} ,
\qquad
g_{L,\mathbf{p}} = \frac{ - \tilde{g}^{\rm intra}_{L,\mathbf{p}} +\tilde{g}^{\rm inter}_{L,\mathbf{p}}( 3\alpha_0^2  -3\alpha_1^2)}{1+3\alpha_0^2+3\alpha_1^2}  ,
\end{equation}
with
\begin{equation}
\tilde{g}^{\rm inter}_{\chi=T,L;\mathbf{p}}= (4\pi\hbar v_F  /3a)\sqrt{\hbar \abs{\mathbf{q}_1}/\rho c_\chi }, \qquad
\tilde{g}^{\rm intra}_{\chi=T,L;\mathbf{p}}=(3 \beta_0 \gamma_0/4) \sqrt{  \hbar\abs{\mathbf{q_1}}/\rho c_{L}  }. \qquad
\end{equation}

For phonons with momentum $\mathbf{p}\approx-\mathbf{q}_2,-\mathbf{q}_3$, the e-phonon scattering matrix can be obtained by the $\mathcal{C}_{3z}$ symmetry as
\begin{equation}    \label{ep C3 trans}
 \hat{g}_{\eta,\chi}(\mathcal{C}_{3z}\mathbf{p})= 
 e^{-i\eta \frac{2\pi}{3} \sigma_z } \left[ \hat{g}_{\eta,\chi}\left( \mathbf{p}\right)
 \right] e^{i\eta \frac{2\pi}{3} \sigma_z}
\end{equation} 
for which $\mathbf{q}_2=\mathcal{C}_{3z}(\mathbf{q}_1)$, $\mathbf{q}_3=\mathcal{C}^{2}_{3z}(\mathbf{q}_1)$ and $\hat{g}_{\eta,\chi}(-\mathbf{p})= 
  \left[ \hat{g}_{\eta,\chi}\left( \mathbf{p} \right) \right]^\dagger$.

Similarly, the e-phonon scattering matrix for valley $-$ can be obtained by the time reversal symmetry  $ \hat{g}_{\eta=-,\chi}(-\mathbf{p})= 
  \left[ \hat{g}_{\eta=+,\chi}\left( \mathbf{p} \right)
 \right]^*$.

\section{Dynamic Kohn anomaly in TBG} 
The nesting of  the Moire Dirac points by phonon wave vector at neutrality  may result in a dynamic Kohn anomaly, for which the phonon self-energy diverges due to e-phonon interaction. As a consequence, the phonon frequency is modified dramatically.
However,  when the electron bands become so flat that the electron velocity $v$ is even smaller than the phonon velocity $s$, the e-phonon scatterings are highly suppressed because of the energy mismatch during the scatterings. For TBG,  when the twist angle approaches the magic angle, the Kohn anomaly may  disappear. We study this phenomenon in the following.  Since we only consider the inter-Moire-Dirac point e-phonon scattering near the two Moire Dirac points in the following text, we use $\mathbf k$ to replace $\tilde{\mathbf k}$ to lighten the notation from now on.

\subsection{Phonon self-energy in TBG}
\begin{figure}    
 \centering 
 \includegraphics[width=8cm]{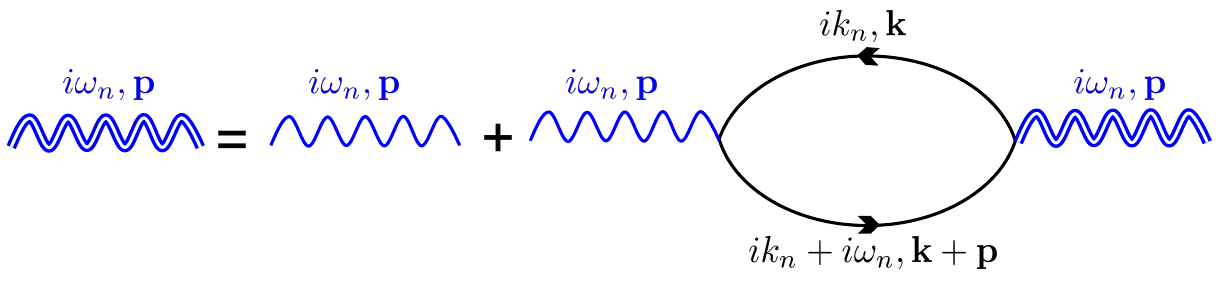}
 \caption{
Feynman diagram of the renormalized phonon propagator. The double wavy line and single wavy line represent the renormalized and bare phonon propagator respectively. The solid line represents the bare electron Green's function.
}\label{fig:phonon_self_energy} 
 \end{figure}

The Feynman diagram of the phonon self-energy is shown in Fig.\ref{fig:phonon_self_energy}, which can be written as 
\begin{equation}
    \Pi_{\chi}(i\omega_n, \mathbf{p})=
 2\sum_{\eta=\pm} \Pi^{\eta}_{\chi}(i\omega_n, \mathbf{p}) 
\end{equation}
where the factor 2 accounts for spin degeneracy, and $\Pi^\eta_{\chi}(i\omega_n, \mathbf{p})$ represents the contributions to the phonon self-energy from the  valley $\eta$, 
\begin{equation}  \begin{aligned}  \label{s-energy form}
    \Pi^{\eta}_{\chi}(i\omega_n, \mathbf{p}) = &
  \frac{1}{\beta}\sum_{i k_{n},\mathbf{k} } 
   \text{Tr}  \Big[
   G^0_{\eta,\mathbf{K}_{M1}}(ik_{n}, \mathbf{k})
    \hat{g}^{\dagger} _{\eta,\chi}(\mathbf{p})   
    G^0_{\eta,\mathbf{K}_{M2}}(ik_{n}+i\omega_n, \mathbf{k}+\mathbf{p}) 
      \hat{g}_{\eta,\chi}(\mathbf{p})   
    \Big],
 \end{aligned}  \end{equation}
with $\beta=1/k_BT$.

The bare electron Green's function near the Moire Dirac point $\mathbf K_{M_i}$ of valley $\eta$ is
\begin{equation}  \label{free e-GF}
    G_{\eta,\mathbf{K}_{Mi}}^{0}(ik_{n}, \mathbf{k})=  
    \frac{ (ik_{n}+\mu)\sigma_0 + v (\mathbf{k}-\mathbf{K}_{Mi})\cdot \boldsymbol{\sigma}_{\eta} }
  {(ik_{n}+\mu)^2-E_{\mathbf{k}-\mathbf{K}_{Mi}}^2 },
 \end{equation} 
where $\boldsymbol{\sigma}_{\eta}\equiv(\eta \sigma_x, -\sigma_y )$, $E_{\mathbf{k}-\mathbf{K}_{Mi}}\equiv v\abs{\mathbf{k}-\mathbf{K}_{Mi}}$ represents the upper band spectrum of Eq.(\ref{eq:effective_H}),  $ik_{n}$ denotes the electronic Matsubara frequency and $\mu$ the chemical potential. 

For the phonon momentum $\mathbf p\approx -\mathbf q_1$, the phonon self-energy after the sum over the electron Matsubara frequency $ik_n$ becomes 
\begin{equation}  \begin{aligned}   \label{eq:Phonon_self_energy}
 & \Pi_{\chi}(i\omega_n, \mathbf{p})= 
    2\sum_{\mathbf{k}}  \sum_{t,t'=\pm} \sum_{\eta=\pm}
   \frac{ n_F(t'  E_{\mathbf{k}-\mathbf{K}_\text{M1}}-\mu) -n_F(tE_{\mathbf{k}+\mathbf{p}-\mathbf{K}_\text{M2}}-\mu) }
   {i\omega_n + t'  E_{\mathbf{k}-\mathbf{K}_{M1}} - t  E_{\mathbf{k}+\mathbf{p}-\mathbf{K}_\text{M2}} }
   \left| \bra{\mathbf{k}+\mathbf{p},t;\eta}\hat{g}_{\eta,\chi}(\mathbf{p})\ket{\mathbf{k},t';\eta}  \right|^2.
\end{aligned} \end{equation} 
Here, $t,t'=\pm1$ represents the two bands of the effective Hamiltonian Eq.(\ref{eq:effective_H}), 
$n_F$ is the Fermi distribution function, 
$\bra{\mathbf{k}+\mathbf{p},t;\eta}\hat{g}_{\eta,\chi}(\mathbf{p})\ket{\mathbf{k},t';\eta}$ is the e-phonon scattering matrix from the electronic state $\mathbf k$ of band $t'$ to the state $\mathbf k+\mathbf p$ of band $t$ in valley $\eta$ caused by the phonon with polarization $\chi$, as shown in Fig.\ref{fig:NCA_diagram}b. For $\chi= L, T$, the e-phonon scattering matrices are respectively
\begin{eqnarray}
  \left| \bra{\mathbf{k}+\mathbf{p},t;\eta}\hat{g}_{\eta,\chi=L}(\mathbf{p})\ket{\mathbf{k},t';\eta}  \right|^2 &=&
  \frac{\abs{g_{L,\mathbf{p}} }^2}{2}  
  \left[     
\alpha_1^2 - t t' \alpha_1^2 \cos{( \varphi_{\mathbf{k}+\mathbf{p}}
+\varphi_{\mathbf{k}} )}      \right], \\
 \left| \bra{\mathbf{k}+\mathbf{p},t;\eta}\hat{g}_{\eta,\chi=T}(\mathbf{p})\ket{\mathbf{k},t';\eta}  \right|^2 &=& \frac{\abs{g_{T,\mathbf{p}} }^2}{2}  
 [(\alpha_0^2+\alpha_1^2)+t t' \alpha_0^2 \cos{( \varphi_{\mathbf{k}+\mathbf{p}} -\varphi_{\mathbf{k}})}  \nonumber\\
 & & +t t' \alpha_1^2 \cos{ ( \varphi_{\mathbf{k}+\mathbf{p}} +\varphi_{\mathbf{k}}) }
 + 2 t'\eta \alpha_0\alpha_1 \cos{ \varphi_{\mathbf{k}} }  + 2 t\eta \alpha_0\alpha_1 \cos{ \varphi_{\mathbf{k}+\mathbf{p}} }],
\end{eqnarray}
where $\varphi_{\mathbf{k}}$ and $\varphi_{\mathbf{k}+\mathbf{p}}$ are the polar angles of the vectors $\mathbf{k}-\mathbf{K}_\text{M1}$ and $\mathbf{k}+\mathbf{p}-\mathbf{K}_\text{M2}$ respectively.

We separate the phonon self-energy in Eq.(\ref{eq:Phonon_self_energy}) to two parts in the calculation: One part corresponds to the contribution at zero temperature and zero doping, which we denote as the vacuum part $\Pi^{\rm vac}_\chi$; the other part corresponds to the contribution at finite temperature and finite doping, which is denoted as the matter part $\Pi^{\rm mat}_\chi$. We then get after the analytic continuation $i\omega_n\to \omega+i 0^+$, 
 \begin{eqnarray}
\Pi^{\text{vac}}_{\chi=T}(\omega, \mathbf{p} ) & = &
 \frac{4\abs{g_{T,\mathbf{p}} }^2}{(2\pi)^2}
  \int d\mathbf{k}  \left( 
        \frac{1}{\omega -E_{ \mathbf{k}-\mathbf{K}_\text{M1} }-E_{\mathbf{k}+\mathbf{p}-\mathbf{K}_\text{M2}} + i0^+ } -\frac{1}{\omega +E_{ \mathbf{k}-\mathbf{K}_\text{M1} } + E_{\mathbf{k}+\mathbf{p}-\mathbf{K}_\text{M2}} + i0^+ }
        \right) \nonumber\\
 && \times
 \frac{1}{2} \left[   (\alpha_0^2+\alpha_1^2)  -\alpha_0^2\cos{(\varphi_{\mathbf{k}+\mathbf{p}}-\varphi_{\mathbf{k}})}
 -\alpha_1^2\cos{(\varphi_{\mathbf{k}+\mathbf{p}}+\varphi_{\mathbf{k}})}
 \right] , \\
\Pi^{\text{vac}}_{\chi=L}(\omega, \mathbf{p} ) & = &
 \frac{4\abs{g_{L,\mathbf{p}} }^2}{(2\pi)^2}
  \int d\mathbf{k}  \left( 
        \frac{1}{\omega -E_{ \mathbf{k}-\mathbf{K}_\text{M1} }-E_{\mathbf{k}+\mathbf{p}-\mathbf{K}_\text{M2}} + i0^+ } -\frac{1}{\omega +E_{ \mathbf{k}-\mathbf{K}_\text{M1} } + E_{\mathbf{k}+\mathbf{p}-\mathbf{K}_\text{M2}} + i0^+ }
        \right) \nonumber\\
 && \times
 \frac{1}{2} \left[   \alpha_1^2  +
\alpha_1^2\cos{(\varphi_{\mathbf{k}+\mathbf{p}}+\varphi_{\mathbf{k}})}
 \right], \\
 \Pi^{\text{mat}}_{\chi}(\omega, \mathbf{p} ) &= &
\Pi_{\chi}(\omega, \mathbf{p} )-\Pi^{\text{vac}}_{\chi}(\omega, \mathbf{p} ) .
\end{eqnarray}

In the case $v\gg s$ and the temperature much less than the bandwidth, the deviation of the total phonon self-energy $\Pi_{\chi}$ from the vacuum part $\Pi^{\text{vac}}_{\chi=T}$ is very small and we can get the information of the Kohn anomaly analytically from $\Pi^{\text{vac}}_{\chi=T}$.
At  zero doping, only inter-band scattering processes can happen, i.e., $tt'=-1$ in Eq.(\ref{eq:Phonon_self_energy}). The vacuum part of the phonon self-energy can be obtained as
\begin{eqnarray}
   &&\text{Re} \left[ \Pi^{\text{vac} }_{\chi=T}(\omega, \mathbf{p}) \right] =
   \abs{g_{T,\mathbf{p}} }^2
    \frac{ \abs{\mathbf{p}+\mathbf{q}_1}   }{ 2 \pi  v}  
   \Bigg\{ -\frac{\alpha_1^2}{4}(\Lambda -1)-\frac{\alpha_1^2}{4}(-\frac{1}{\Lambda }+1) \nonumber\\
  &&\    - \frac{ \left( \alpha_0^2+\alpha_1^2 ( \omega^2/ v^2 |\mathbf{p}+\mathbf{q}_1|^2 \right) } {\sqrt{ 1- \omega^2/v^2 |\mathbf{p}+\mathbf{q}_1|^2 }}   \sum_{t=\pm}
 \left[\arctan( \frac{ \Lambda +t  \omega/v |\mathbf{p}+\mathbf{q}_1|  }{\sqrt{ 1- \omega^2/v^2 |\mathbf{p}+\mathbf{q}_1|^2 }})-\arctan( \frac{ 1 + t\omega/v|\mathbf{p}+\mathbf{q}_1|   }{\sqrt{ 1-( \omega)^2/v^2|\mathbf{p}+\mathbf{q}_1|^2 }})    \right]
  \Bigg\}, \label{eq:SE_vac_T}\\
  && \text{Re} \left[ \Pi^{\text{vac} }_{\chi=L}(\omega, \mathbf{p}) \right]=
 \abs{g_{L,\mathbf{p}} }^2
    \frac{  \abs{\mathbf{p}+\mathbf{q}_1}   }{2 \pi v} 
    \Bigg\{ -\frac{\alpha_1^2}{2}\Lambda+\frac{\alpha_1^2}{2}\frac{1}{\Lambda} \nonumber\\
      && +\alpha_1^2 \frac{ \left( 1+\omega^2/ v^2 |\mathbf{p}+\mathbf{q}_1|^2 \right) } {\sqrt{ 1- \omega^2/v^2 |\mathbf{p}+\mathbf{q}_1|^2 }}   \sum_{t=\pm}
  \left[\arctan( \frac{ \Lambda +t  \omega/v |\mathbf{p}+\mathbf{q}_1|  }{\sqrt{ 1- \omega^2/v^2 |\mathbf{p}+\mathbf{q}_1|^2 }})-\arctan( \frac{ 1 + t\omega/v|\mathbf{p}+\mathbf{q}_1|   }{\sqrt{ 1-( \omega)^2/v^2|\mathbf{p}+\mathbf{q}_1|^2 }})    \right]   \Bigg\} \label{eq:SE_vac_L}
 \end{eqnarray}
 and 
 \begin{eqnarray}
 \text{Im} \left[ \Pi^{\text{vac} }_{\chi=T}( \omega, \mathbf{p}) \right] &=&
\abs{g_{T,\mathbf{p}} }^2
   \frac{ \abs{\mathbf{p}+\mathbf{q}_1}   }{ 4 v } 
   \frac{ \left( \alpha_0^2+\alpha_1^2 (\omega)^2/E_{\mathbf{p}+\mathbf{q}_1}^2  \right) }{ \sqrt{ \abs{( \omega)^2/E_{\mathbf{p}+\mathbf{q}_1}^2 -1}} }  
  \left[ -\Theta(\omega-E_{\mathbf{p}+\mathbf{q}_1})  + \Theta(-\omega-E_{\mathbf{p}+\mathbf{q}_1})  \right], \label{eq:Im_Pi_T}\\
  \text{Im} \, \left[ \Pi^{\text{vac} }_{\chi=L}(\omega, \mathbf{p}) \right] &=&
 \abs{g_{L,\mathbf{p}} }^2
 \alpha_1^2
    \frac{ \abs{\mathbf{p}+\mathbf{q}_1}   }{ 4 v } 
    \sqrt{\abs{ \frac{\omega^2}{v^2 \abs{\mathbf{p}+\mathbf{q}_1}^2}-1 } }
  \left[ -\Theta(\omega-v \abs{\mathbf{p}+\mathbf{q}_1} ) + \Theta(-\omega-v \abs{\mathbf{p}+\mathbf{q}_1} )   \right]  \label{eq:Im_Pi_L}
\end{eqnarray}
where 
$\Lambda=1+\frac{2 E_c}{v\abs{\mathbf{p}+\mathbf{q}_1}}+
  \sqrt{\Big( 1+\frac{2E_c}{v\abs{\mathbf{p}+\mathbf{q}_1}} \Big)^2+1}$
with $E_c\sim v q_1$ the high energy cutoff of the Moire electrons, and $\Theta(\omega-E_{\mathbf{p}+\mathbf{q}_1})$ is the Heaviside step function.

The matter part of the phonon self-energy comes from the contribution of finite temperature and finite doping.  For the $\rm TA$ mode, it is
\begin{eqnarray}
\Pi^{\text{mat}}_{\chi=T}(i\omega_n, \mathbf{p} )  &= &
 \frac{4\abs{g_{T,\mathbf{p}} }^2}{(2\pi)^2}
  \int d\mathbf{k}  
 \sum_{t, t'=\pm 1}  \frac{ n_F(t'  E_{\mathbf{k}-\mathbf{K}_\text{M1}}-\mu) -n_F(tE_{\mathbf{k}+\mathbf{p}-\mathbf{K}_\text{M2}}-\mu)-\delta_{t,-t'}(\delta_{t',-1}-\delta_{t,-1}) }
   {i\omega_n + t'  E_{\mathbf{k}-\mathbf{K}_\text{M1}} - t  E_{\mathbf{k}+\mathbf{p}-\mathbf{K}_\text{M2}} } \nonumber\\
   &&\frac{1}{2}   \left[ 
 (\alpha_0^2+\alpha_1^2)+t t' \alpha_0^2 \cos{( \varphi_{\mathbf{k}+\mathbf{p}} -\varphi_{\mathbf{k}})} +t t' \alpha_1^2 \cos{ ( \varphi_{\mathbf{k}+\mathbf{p}} +\varphi_{\mathbf{k}}) }    \right]. 
 \end{eqnarray}

Since $1-n_F(-E_{\mathbf{k}-\mathbf{K}_\text{M1}}-\mu)=n_F(E_{\mathbf{k}-\mathbf{K}_\text{M1}}+\mu)$, the above equation can be simplified as 
\begin{equation} \label{eq:Pi_mat_T}
 \Pi^{\text{mat}}_{\chi=T}  (i\omega_n,  \mathbf{p} ) = 
  \frac{4\abs{g_{T,\mathbf{p}} }^2}{(2\pi)^2}
   \int_{0}^{\infty}  \frac{d( E_{\mathbf{k}-\mathbf{K}_\text{M1}} )}{v^2} 
 \Big(    n_F( E_{\mathbf{k}-\mathbf{K}_\text{M1}}+\mu )+n_F( E_{\mathbf{k}-\mathbf{K}_\text{M1}}-\mu )  \Big)
\left[ \sum_{t=\pm} \int_{0}^{2\pi} d\varphi_{\mathbf{k}} 
  \, F_{t,\chi=T}( \mathbf{k},\mathbf{p}, i\omega_n )
  \right]   
 \end{equation}
where 
\begin{eqnarray}
F_{t,\chi=T}( i\omega_n, \mathbf{k},\mathbf{p} ) &=&   
    \frac{ 1 }
{  \big(E_{\mathbf{k}-\mathbf{K}_\text{M1}}+t (i\omega_n)\big)^2-E_{\mathbf{k}+\mathbf{p}-\mathbf{K}_\text{M2}}^2  } 
\Big[  (\alpha_0^2+\alpha_1^2) \big( E_{\mathbf{k}-\mathbf{K}_\text{M1}}+t (i\omega_n) \big) E_{\mathbf{k}-\mathbf{K}_\text{M1}}     \nonumber\\
&+& \alpha_0^2E_{\mathbf{k}-\mathbf{K}_\text{M1}}E_{\mathbf{k}+\mathbf{p}-\mathbf{K}_\text{M2}}\cos{(\varphi_{\mathbf{k}+\mathbf{p}}-\varphi_{\mathbf{k}})}
+\alpha_1^2E_{\mathbf{k}-\mathbf{K}_\text{M1}}E_{\mathbf{k}+\mathbf{p}-\mathbf{K}_\text{M2}}\cos{(\varphi_{\mathbf{k}+\mathbf{p}}+\varphi_{\mathbf{k}})}    \Big], 
\end{eqnarray}
and the integration over the angel $\varphi_{\mathbf k}$ can be worked out as
\begin{eqnarray}
&& \Phi_T(i\omega_n, \mathbf p, E_{\mathbf k-\mathbf k_{M1}})\equiv \sum_{t=\pm} \int_{0}^{2\pi} d\varphi_{\mathbf{k}} 
  \, F_{t,\chi=T} ( \mathbf{k},\mathbf{p}, i\omega_n )  =
  2\pi \Big[ \alpha_1^2\frac{ (i\omega_n)^2 }{ v^2|\mathbf{p}+\mathbf{q}_1|^2 }+\alpha_0^2 \Big] \sum_{t=\pm}
 \bigg\{  
    -\frac{1}{2}   \nonumber\\
  && 
    +   \frac{
    2E_{\mathbf{k}-\mathbf{K}_\text{M1}}^2+ 2t (i\omega_n)\,E_{\mathbf{k}-\mathbf{K}_\text{M1}}+\frac{1}{2}
  \left( (i\omega_n)^2 - v^2 |\mathbf{p}+\mathbf{q}_1|^2 \right) 
  }{
  \sqrt{ 4\left((i\omega_n)^2-v^2|\mathbf{p}+\mathbf{q}_1|^2\right)E_{\mathbf{k}-\mathbf{K}_\text{M1}}^2 +
    4t(i\omega_n)\left((i\omega_n)^2-v^2 |\mathbf{p}+\mathbf{q}_1|^2\right)E_{\mathbf{k}-\mathbf{K}_\text{M1}}
    +\left((i\omega_n)^2-v^2|\mathbf{p}+\mathbf{q}_1|^2\right)^2 }
  }
 \bigg\} .
\end{eqnarray}

Similarly, for the $\rm LA$ mode, the matter part of the phonon self-energy is 
    \begin{eqnarray} \label{eq:Pi_mat_L}
      \Pi^{\text{mat}}_{\chi=L}(i\omega_n, \mathbf{p} )  &= &
       \frac{4\abs{g_{T,\mathbf{p}} }^2}{(2\pi)^2}
        \int d\mathbf{k}  
       \sum_{t, t'=\pm 1}  \frac{ n_F(t'  E_{\mathbf{k}-\mathbf{K}_\text{M1}}-\mu) -n_F(tE_{\mathbf{k}+\mathbf{p}-\mathbf{K}_\text{M2}}-\mu)-\delta_{t,-t'}(\delta_{t',-1}-\delta_{t,-1}) }
         {i\omega_n + t'  E_{\mathbf{k}-\mathbf{K}_\text{M1}} - t  E_{\mathbf{k}+\mathbf{p}-\mathbf{K}_\text{M2}} } \nonumber\\
         && \quad \quad \quad \quad \quad \quad \frac{1}{2}  
        \left[     
      \alpha_1^2 - t t' \alpha_1^2 \cos{( \varphi_{\mathbf{k}+\mathbf{p}}
      +\varphi_{\mathbf{k}} )}      \right] \nonumber \\
      &=&
        \frac{4\abs{g_{T,\mathbf{p}} }^2}{(2\pi)^2}
         \int_{0}^{\infty}  \frac{d( E_{\mathbf{k}-\mathbf{K}_\text{M1}} )}{v^2} 
       \Big(    n_F( E_{\mathbf{k}-\mathbf{K}_\text{M1}}+\mu )+n_F( E_{\mathbf{k}-\mathbf{K}_\text{M1}}-\mu )  \Big)
      \left[ \sum_{t=\pm} \int_{0}^{2\pi} d\varphi_{\mathbf{k}} 
        \, F_{t,\chi=L}( i\omega_n, \mathbf{k},\mathbf{p})
        \right]  \nonumber\\
       \end{eqnarray}
       where
       \begin{eqnarray} 
      && F_{t,\chi=L} ( i\omega_n, \mathbf{k},\mathbf{p})= \nonumber\\
      &&\quad \quad \frac{\alpha_1^2}{\big(E_{\mathbf{k}-\mathbf{K}_\text{M1}}+t (i\omega_n)\big)^2-E_{\mathbf{k}+\mathbf{p}-\mathbf{K}_\text{M2}}^2}
        \Big[  \big( E_{\mathbf{k}-\mathbf{K}_\text{M1}}+t (i\omega_n) \big) E_{\mathbf{k}-\mathbf{K}_\text{M1}}-
      E_{\mathbf{k}-\mathbf{K}_\text{M1}}E_{\mathbf{k}+\mathbf{p}-\mathbf{K}_\text{M2}}\cos{(\varphi_{\mathbf{k}+\mathbf{p}}+\varphi_{\mathbf{k}})}
        \Big]  
       \end{eqnarray}
       and
      \begin{equation} \begin{aligned}
     & \Phi_L(i\omega_n, \mathbf p, E_{\mathbf k-\mathbf k_{M1}})\equiv \sum_{t=\pm 1}  \int_{0}^{2\pi} d\varphi_{\mathbf{k}}   \, F_{t,\chi=L} ( i\omega_n, \mathbf{k},\mathbf{p})= 2\pi\alpha_1^2 \frac{(i\omega_n)^2}{v^2|\mathbf{p}+\mathbf{q}_1|^2} \sum_{t=\pm 1}
            \\  & 
          \left[ \frac{1}{2} +   \frac{ \left(v^2|\mathbf{p}+\mathbf{q}_1|^2/(i\omega_n)^2-1\right)
          \left( 2E_{\mathbf{k}-\mathbf{K}_\text{M1}}^2+ 2t (i\omega_n)\,E_{\mathbf{k}-\mathbf{K}_\text{M1}}+\frac{1}{2}
         (i\omega_n)^2  \right)
        }{
        \sqrt{ 4\left((i\omega_n)^2-v^2|\mathbf{p}+\mathbf{q}_1|^2\right)E_{\mathbf{k}-\mathbf{K}_\text{M1}}^2 +
          4t(i\omega_n)\left((i\omega_n)^2-v^2 |\mathbf{p}+\mathbf{q}_1|^2\right)E_{\mathbf{k}-\mathbf{K}_\text{M1}}
          +\left((i\omega_n)^2-v^2|\mathbf{p}+\mathbf{q}_1|^2\right)^2 }
        }
       \right] .
      \end{aligned} \end{equation}

At finite temperature, it is hard to get an analytic result of the integration in Eq.(\ref{eq:Pi_mat_T}) and (\ref{eq:Pi_mat_L}). We then evaluate $\Pi^{\rm mat}_\chi$ numerically and add it to the total phonon self-energy $\Pi_{\chi}$.

\subsection{Dynamic Kohn anomaly for the TA mode}

We first consider the case with zero doping. The renormalized phonon frequency due to the e-phonon interaction can be obtained from the pole of the renormalized phonon GF,
$D_\chi(i\omega_n, \mathbf p)=D^0_\chi(i\omega_n, \mathbf p)/[1-D^0_\chi(i\omega_n, \mathbf p)
\Pi_\chi(i\omega_n, \mathbf p)]$, where $D^0_\chi(i\omega_n, \mathbf p)=2\omega_{0, \mathbf p, \chi}/[(i\omega_n)^2-\omega^2_{0, \mathbf p, \chi}]$ is the bare phonon propagator. The renormalized phonon frequency $\omega_{\mathbf p, \chi}$ then satisfies
\begin{equation}
\omega_{\mathbf p, \chi}^2/\omega^2_{0, \mathbf p, \chi}-1=2\rm Re[\Pi_\chi(\omega, \mathbf p)]/\omega_{0, \mathbf p, \chi},
\end{equation}
where the phonon self-energy $\Pi_\chi(\omega, \mathbf p)$ is worked out in the last subsection. 

Figure.\ref{fig:TA_self_energy}(a)-(e) and Fig.\ref{fig:LA_self_energy}(a)-(e) show the real part of $\Pi^{\rm vac}_\chi$ for the TA and LA mode respectively.
From Eq.(\ref{eq:SE_vac_T}) and (\ref{eq:SE_vac_L}), the real part of $\Pi^{\rm vac}_\chi$ for both the TA and LA phonon diverges at 
\begin{equation}\label{eq:KA_condition}
\omega=v |\mathbf p+\mathbf q_1|.
\end{equation}
In this condition, the imaginary part of the self-energy for the TA mode  in Eq.(\ref{eq:Im_Pi_T}) diverges too, whereas the imaginary of the self-energy for the LA mode is finite. For the twist angle away from the magic angle, the electron velocity $v$ is much greater than the phonon velocity $s$ and the phonon frequency $\omega$ is negligible in Eq.(\ref{eq:KA_condition}). The Kohn anomaly, i.e., the divergence of the self-energy then occurs at $\mathbf p\approx -\mathbf q_1$ due to the nesting of the Fermi points at zero doping.

For twist angle close to the magic angle, the two bands of the Hamiltonan Eq.(\ref{eq:effective_H})  becomes very narrow. In this case, the phonon energy $\omega_0=s p$ is not negligible compared to the electron energy and the dynamics of the phonon  becomes important. For the perturbative self-energy $\Pi_\chi(\omega, \mathbf p)$, we can set $\omega=\omega_0$. The divergence of the self-energy then occurs at phonon momentum $\mathbf p$ that satisfies
\begin{equation}
s  p=v|\mathbf p+\mathbf q_1|.
\end{equation}
The momentum mismatch of the nesting condition $\delta \mathbf q\equiv \mathbf p+\mathbf q_1$  is compensated by the dynamics of the phonon.

\begin{figure}
  \includegraphics[width=18cm]{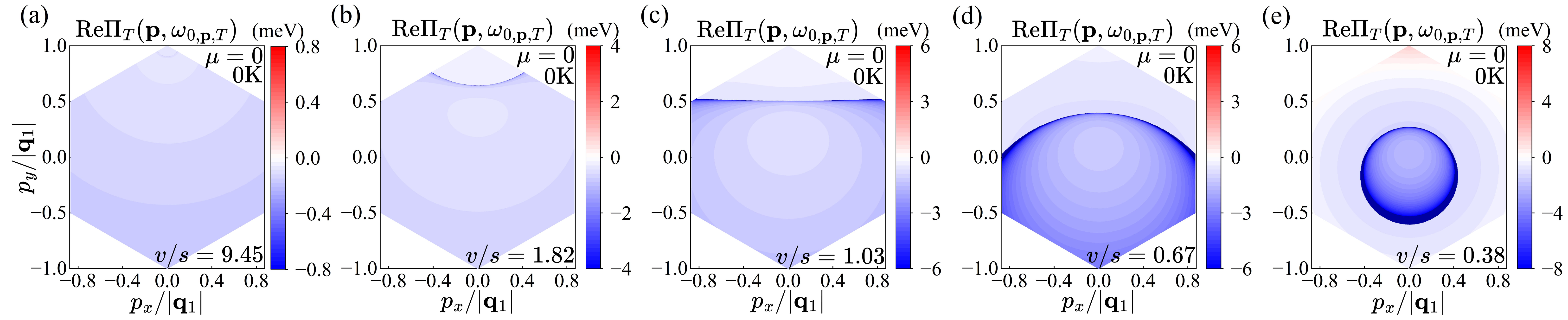}
      \includegraphics[width=18cm]{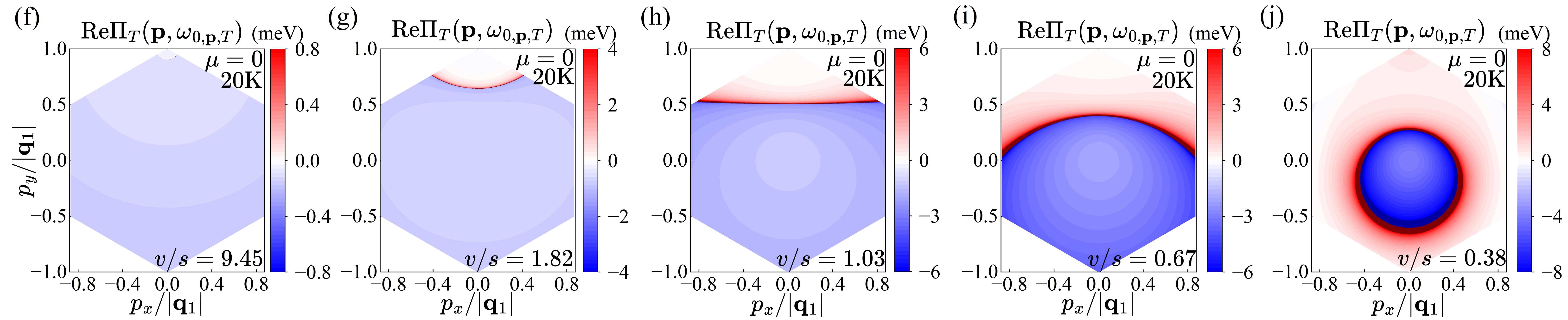}
 \caption{The real part of the TA phonon self-energy at zero doping: (a)-(e) at zero temperature and (f)-(j) at $T=20K$ for different $v/s$.  }\label{fig:TA_self_energy}
\end{figure}

The orbit of the phonon momentum $\mathbf p$ satisfying Eq.(\ref{eq:KA_condition}) is a circle in momentum space with the center $(p_x, p_y)=(0, \frac{v^2}{v^2-s^2}q_1)$ and radius $\frac{sv}{|s^2-v^2|}q_1$ for $s\neq v$ and a straight line for $s=v$.
This is clearly seen from the divergence lines in Fig.\ref{fig:TA_self_energy}(a)-(e) for the vacuum part of the TA self-energy with different $v/s$.
Figure \ref{fig:TA_SE_cut}(a) shows the cut of the TA phonon self-energy along the line $p_x=0$ from Fig.\ref{fig:TA_self_energy}(a)-(e). It is clear that with the decrease of $v/s$, the phonon momentum  where the divergence of the self-energy occurs moves from $p_y=q_1$ towards $p_y=0$.

\begin{figure}
  \includegraphics[width=16cm]{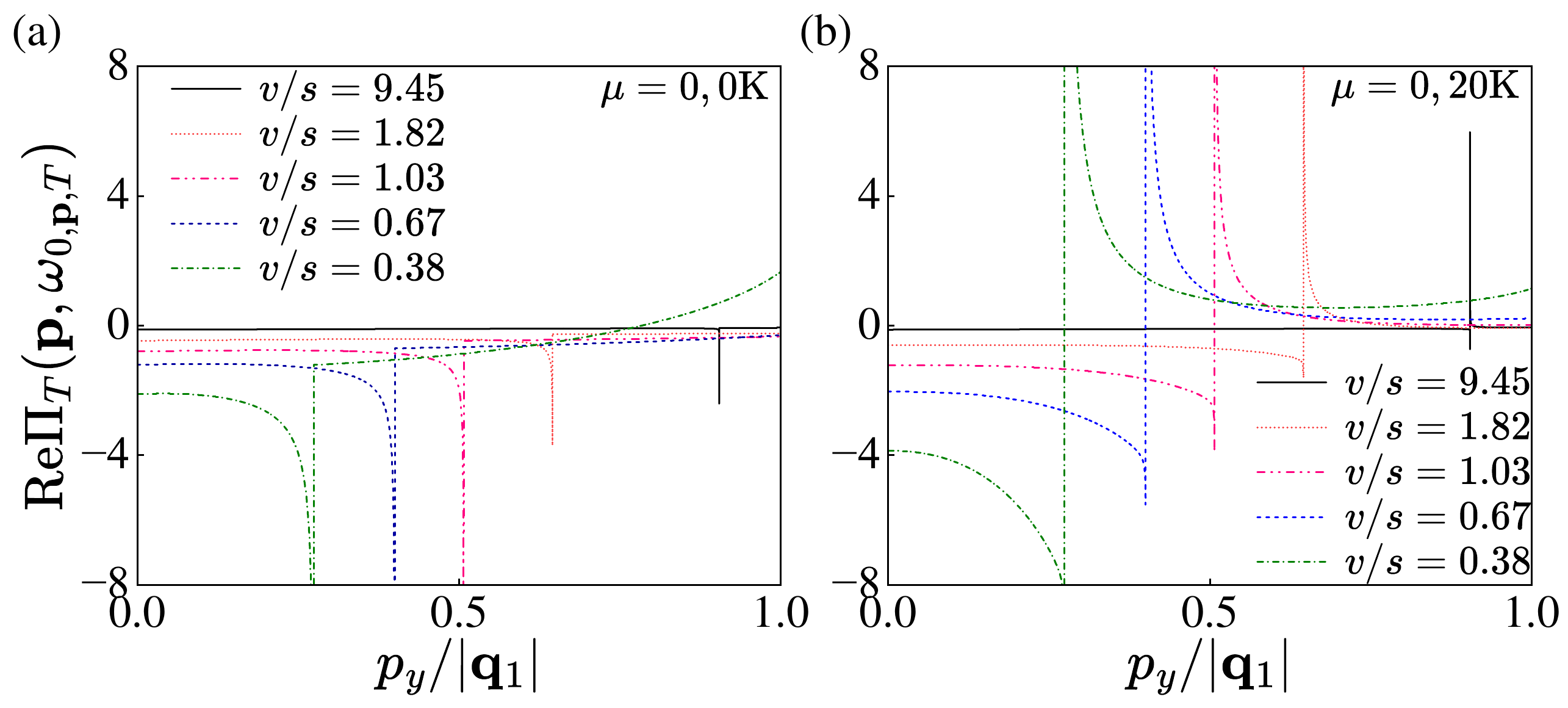}
 \caption{ The real part of the TA phonon self-energy at (a)$T=0K$ and (b)$T=20K$ along the line $p_x=0$ for different $v/s$.  }\label{fig:TA_SE_cut}
\end{figure}

We should note that since
 we have set the e-phonon coupling constant $g_\chi(\mathbf p) \approx g_\chi(-\mathbf q_1)$ in the calculation, the result of the self-energy in Eq.(\ref{eq:SE_vac_T}) and (\ref{eq:SE_vac_L}) is quantitatively accurate only near $\mathbf p\approx -\mathbf q_1$. 
 However, the divergence condition in Eq.(\ref{eq:KA_condition}) comes from the energy conservation 
 and should still be valid for $\mathbf p$ deviates from $-\mathbf q_1$.
 A particularly interesting question is whether the dynamic Kohn anomaly happens when the electron bands are  flat. Here we can see that for the flat bands, i.e., $v/s=0$, the dynamic Kohn anomaly could only happen at phonon momentum $\mathbf p=0$ because in this case the e-phonon scattering is elastic. However, for the acoustic mode, either TA or LA mode, the bare phonon frequency at $\mathbf p=0$ is already zero. So the phonon softening can no longer happen, i.e., the dynamic  Kohn anomaly of the acoustic modes does not occur for the completely flat electron bands.

We next study the effect of the matter part due to the finite temperature. For $T$ much less than the electron bandwidth $\Lambda\sim v q_1$, the correction to the phonon self-energy by the finite temperature is very small, i.e., we can neglect the matter part of the phonon self-energy
and only include the vacuum part. However, for TBG, the electron bandwidth can  become very small by tuning the twist angle. When the bandwidth is comparable to the temperature, the contribution from the matter part becomes significant and one needs to include this part in the phonon self energy. 
 It is hard to get an analytic result for the matter part of the self-energy. For the reason,  we compute the matter part of the TA phonon self-energy numerically and show the TA phonon self-energy including the matter part at $T=20K$  in Fig.\ref{fig:TA_self_energy}(f)-(j) for different $v/s$. The cut of the phonon self-energy  at $T=20K$ along the line $p_x=0$ is shown in Fig.\ref{fig:TA_SE_cut}(b). It turns out that the matter part of the TA phonon mode has opposite sign as the vacuum part.  From Fig.\ref{fig:TA_self_energy} and Fig.\ref{fig:TA_SE_cut}, we can see that the finite temperature does not change the position of the phonon momentum where the Kohn anomaly occurs. However, it may change the value of the phonon self-energy significantly. Particularly, when the temperature is comparable to the electron bandwidth,  the phonon self-energy may change sign at the momenta where the dynamic Kohn anomaly occurs, 
 which results in a change from phonon softening to phonon hardening for momenta on the two sides of the Kohn anomaly, as shown in Fig.\ref{fig:TA_self_energy} and Fig.\ref{fig:TA_SE_cut}.

\subsection{Dynamic Kohn anomaly for the LA mode}
The self-energy of the LA phonon mode at zero temperature is shown in Fig.\ref{fig:LA_self_energy} for different $v/s$. Different from the TA mode, the peak of the LA phonon self-energy near the Kohn anomaly can be positive for the LA mode even at zero temperature, as can be also seen from the cut of self-energy along the line $p_x=0$ in Fig.\ref{fig:LA_SE_cut}a. On the other hand, the matter part of the LA phonon self-energy is small and positive. It vanishes at the Kohn anomaly instead of diverging. For the reason, the finite temperature does not change the LA phonon self-energy qualitatively, as shown in Fig.\ref{fig:LA_self_energy}  and Fig.\ref{fig:LA_SE_cut}.

\begin{figure}
  \includegraphics[width=18cm]{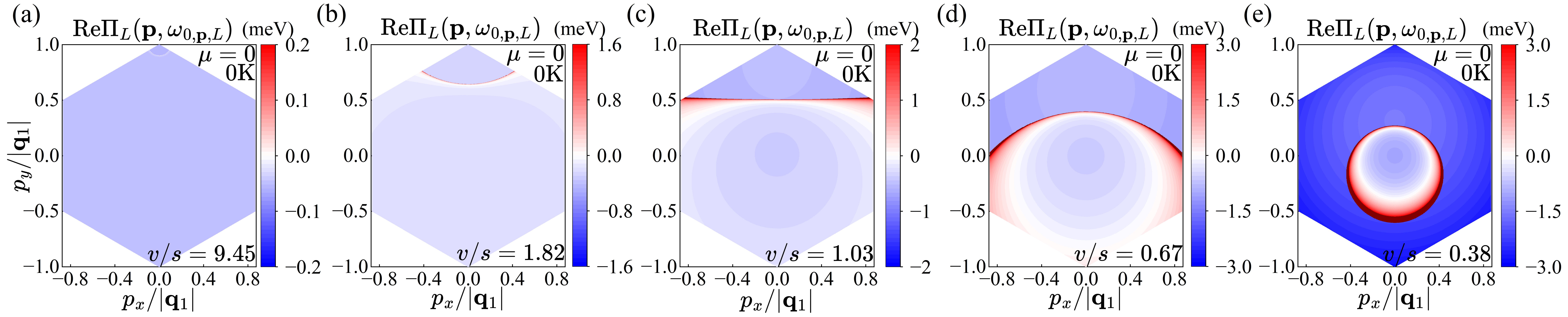}
    \includegraphics[width=18cm]{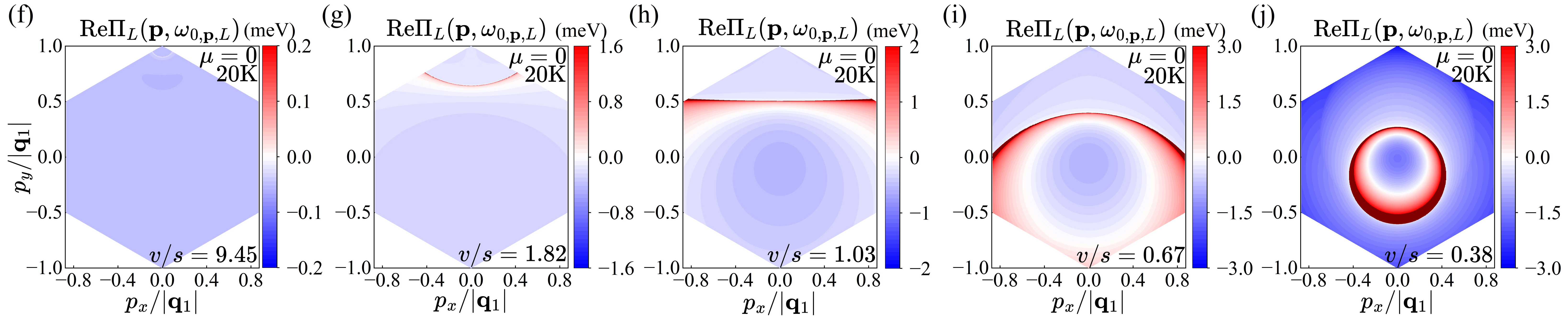}
 \caption{ The real part of the LA phonon self-energy: (a)-(e) at zero temperature and (f)-(j) at $T=20K$ for different $v/s$. }\label{fig:LA_self_energy}
\end{figure}

\begin{figure}
  \includegraphics[width=16cm]{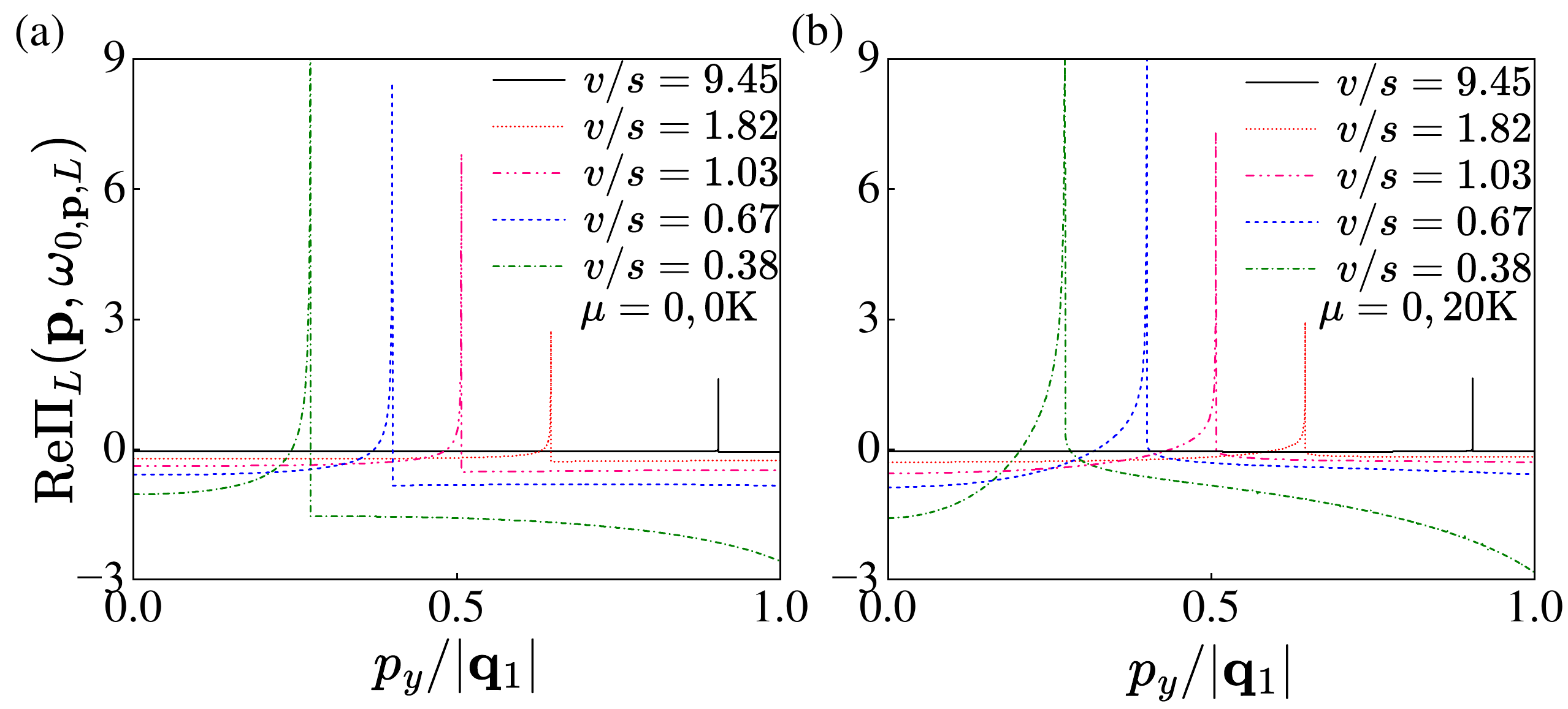}
 \caption{The real part of the phonon self-energy for the LA  mode along the line $p_x=0$ in the BZ (a) at $T=0$ and (b) at $T=20K$. }\label{fig:LA_SE_cut}
\end{figure}

\end{widetext}

\subsection{Dynamic Kohn anomaly at finite  doping}
\begin{figure}
  \includegraphics[width=18cm]{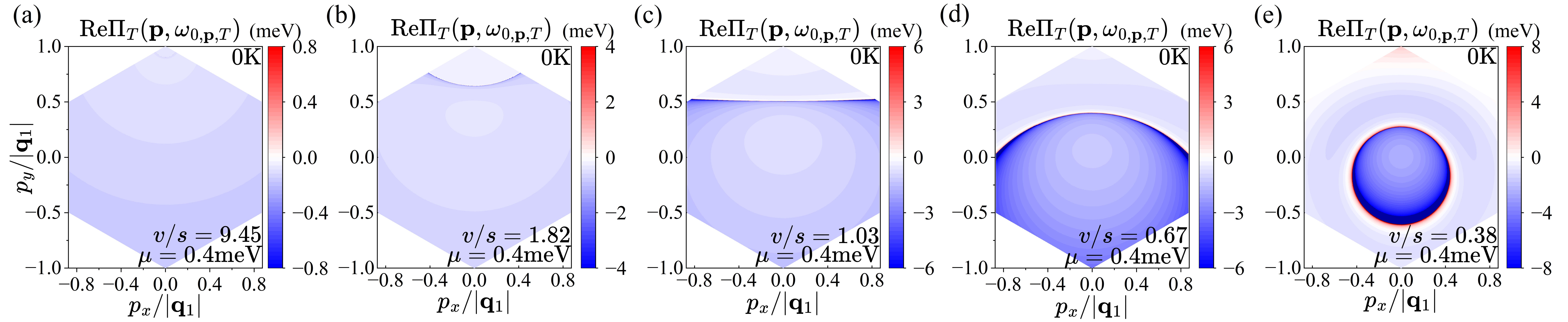}
    \includegraphics[width=18cm]{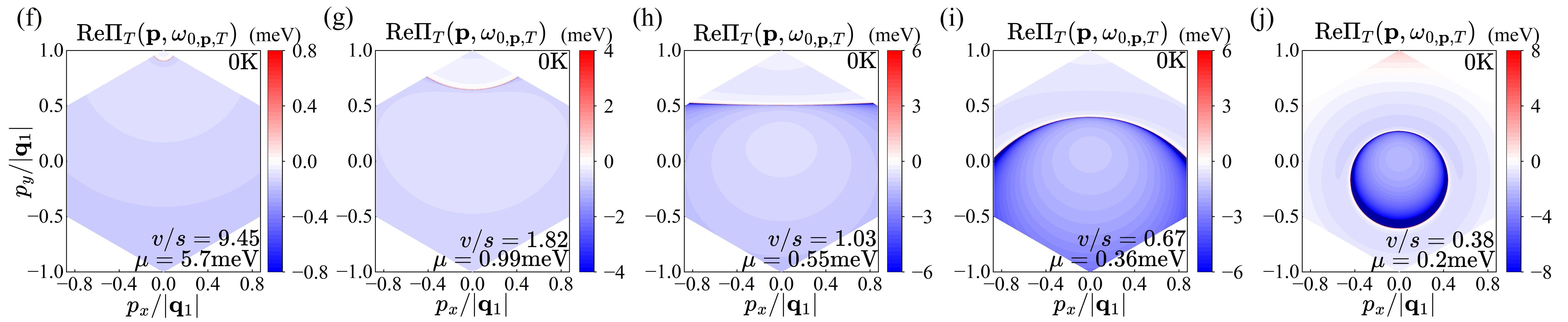}
 \caption{(a)-(e) The self-energy of the TA mode at zero temperature and same finite chemical potential $\mu=0.4 \rm meV$ for different $v/s$. (f)-(j)The self-energy of the TA mode at zero temperature and same filling factor $\nu=0.75$ for different $v/s$.}\label{fig:TA_doping}
\end{figure}

The above results are obtained at zero doping, i.e., the electron Fermi energy crosses the Moire Dirac points. In this subsection, we discuss the dynamic Kohn anomaly at finite doping. For simplicity, we only show the phonon self-energy of the TA mode at finite doping and zero temperature. For the same doping at two Moire Dirac points, the TA phonon self-energy can be obtained by setting $\mu$ finite in Eq.(\ref{eq:Pi_mat_T}). In this case the divergence condition of the TA phonon self-energy is the same as for the zero doping. We show the plot of the TA  phonon self-energy at finite doping in Fig.\ref{fig:TA_doping}. We can see that the finite doping may change the sign of the TA phonon self-energy near the Kohn anomaly. This effect is similar to that of  finite temperature because both doping and finite temperature may activate the electrons at energy $E>0$ and  result in a positive matter part of the TA phonon self-energy near the Kohn anomaly.

\section{Discussions and conclusion}

The dynamic Kohn anomaly in TBG is different from the Kohn anomaly observed in ordinary metal, for which a cusp in the phonon dispersion appears at $q=2k_F$. The difference originates from the relativistic chiral band structure of the TBG we employed in the continuum model. Similar band structure also applies for graphene and results in a novel dynamic Kohn anomaly in such system, as studied in Ref.\cite{Graphene2008}. For both the graphene and undoped TBG, the dynamics of the phonon can not be neglected. This is revealed in the dependence  on the phonon frequency of the phonon wave vector for which the Kohn anomaly occurs. 
However, the dynamic Kohn anomaly in graphene in Ref.\cite{Graphene2008} is also different from that in TBG we obtained in this work.
The former comes from the intravalley e-phonon scatterings in graphene, whereas the latter comes from the inter-valley electron scatterings mediated by phonons in TBG. For the reason, the phonon wave-vectors where the dynamic Kohn anomaly occurs are different in the two systems. For graphene, the Kohn anomaly occurs at $q=\omega/v$ or $2k_F-\omega/q$ for the TO and LO phonons. 

Another system which has similar dynamic Kohn anomaly as TBG is the WSM. The Kohn anomaly due to inter-Weyl-node electron scatterings by phonons in tantalum phosphide (TaP) has been studied both theoretically and experimentally in Ref.\cite{Nguyen2020} The phonon momenta where the dynamic Kohn anomaly occurs in TaP satisfy the same condition as for TBG, i.e., Eq.(\ref{eq:KA_condition}).
However, since the TaP WSM is three dimensional,  the power of the divergence at the Kohn anomaly is different in the two systems. Without screening, the phonon self-energy in TaP diverges logarithmically at the Kohn anomaly, whereas in TBG, the self energy diverges as a power law as shown in Eq.(\ref{eq:SE_vac_T}) and (\ref{eq:SE_vac_L}).  Besides, in WSM, the electron velocity is usually much greater than the phonon velocity. For the reason, the phonon wave-vector for which the Kohn anomaly  occurs is very close to the nesting momentum, i.e., the distance of the two Weyl nodes. However, in TBG, by tuning the twisting angle, the electron velocity may become smaller than the phonon velocity. In this case, the phonon momenta for which  the dynamic Kohn anomaly  occurs deviate significantly from the nesting momentum or the distance of the two Moire Dirac points. At magic angle when the  two electron bands crossing the Fermi energy become almost flat, the dynamic Kohn anomaly disappears  because the inter-valley e-phonon scattering cannot satisfy the energy and momentum conservation at the same time.

The Kohn anomaly can be observed through inelastic x-ray and neutron scattering, as done on WSM TaP. The softening or hardening of the phonon frequency may result in modification of the superconductivity or other phonon-related properties in the TBG, which we will leave for a future study. 

In conclusion, we studied the e-phonon interaction of the  acoustic phonon modes with the electrons in an effective  two band system in the TBG under the continuum model. Based on this model, we studied the effect of the e-phonon interaction on the phonon frequency, i.e., the dynamic Kohn anomaly. We show that the Kohn anomaly in the TBG has different properties from that in ordinary metal and other relativistic materials, such as graphene and WSM. Moreover, by tuning the twist angle, the dynamic Kohn anomaly in TBG is also tunable. Especially, at magic angle when the two bands crossing the Fermi energy becomes almost flat, the dynamic Kohn anomaly in TBG disappears. The Kohn anomaly in TBG may be detected by inelastic x-ray or neutron scatterings.

\section{Acknowledgement}
This work is supported by the National NSF of China under Grant No. 11974166 and the NSF of Jiangsu Province under Grant No.BK20231398.

\appendix


\begin{thebibliography}{99}

\bibitem{Cao2018}Cao Y, Fatemi V, Fang S, Watanabe K, Taniguchi T, Kaxiras E and Jarillo-Herrero P 2018 {\it Nature} 
\href{https://doi.org/10.1038/nature26160}{{\bf556} 43–50} 

\bibitem{Po2018}Po H C, Zou L, Vishwanath A and Senthil T 2018 {\it Phys. Rev.} X 
\href{https://doi.org/10.1103/PhysRevX.8.031089}{{\bf8}, 031089}
 
\bibitem{XLu2019}Lu X, Stepanov P, Yang W, Xie M, Aamir M A, Das I, Urgell C, Watanabe K, Taniguchi T, Zhang G, Bachtold A, MacDonald A H and Efetov D K 2019 {\it Nature} 
\href{https://doi.org/10.1038/s41586-019-1695-0}{{\bf574} 653–7}

\bibitem{Yankowitz_2019}Yankowitz M, Chen S, Polshyn H, Zhang Y, Watanabe K, Taniguchi T, Graf D, Young A F and Dean C R 2019 {\it Science} 
\href{https://www.science.org/doi/10.1126/science.aav1910}{{\bf363} 1059–64}

\bibitem{Stepanov2020} Stepanov P, Das I, Lu X, Fahimniya A, Watanabe K, Taniguchi T, Koppens F H L, Lischner J, Levitov L and Efetov D K 2020 {\it Nature} 
\href{https://doi.org/10.1038/s41586-020-2459-6}{{\bf583} 375–8}

\bibitem{Saito2020}Saito Y, Ge J, Watanabe K, Taniguchi T and Young A F 2020 {\it Nat. Phys.} 
\href{https://doi.org/10.1038/s41567-020-0928-3}{{\bf16} 926–30}

\bibitem{Nuckolls2023}Nuckolls K P, Lee R L, Oh M, Wong D, Soejima T, Hong J P, Călugăru D, Herzog-Arbeitman J, Bernevig B A, Watanabe K, Taniguchi T, Regnault N, Zaletel M P and Yazdani A 2023 {\it Nature} 
\href{https://doi.org/10.1038/s41586-023-06226-x}{{\bf620} 525–32} 

\bibitem{Ledwith_2020}Ledwith P J, Tarnopolsky G, Khalaf E and Vishwanath A 2020 {\it Phys. Rev. Res.} 
\href{https://doi.org/10.1103/PhysRevResearch.2.023237}{{\bf2} 023237}

\bibitem{Repellin_2020}Repellin C and Senthil T 2020 {\it Phys. Rev. Res.} 
\href{https://doi.org/10.1103/PhysRevResearch.2.023238}{{\bf2} 023238}

\bibitem{Wilhelm_2021}Wilhelm P, Lang T C and Läuchli A M 2021 {\it Phys. Rev.} B 
\href{https://doi.org/10.1103/PhysRevB.103.125406}{{\bf103} 125406}

\bibitem{Xie_2021}Xie Y, Pierce A T, Park J M, Parker D E, Khalaf E, Ledwith P, Cao Y, Lee S H, Chen S, Forrester P R, Watanabe K, Taniguchi T, Vishwanath A, Jarillo-Herrero P and Yacoby A 2021 {\it Nature} 
\href{https://doi.org/10.1038/s41586-021-04002-3}{{\bf600} 439–43}

\bibitem{Stepanov_2021}Stepanov P, Xie M, Taniguchi T, Watanabe K, Lu X, MacDonald A H, Bernevig B A and Efetov D K 2021 {\it Phys. Rev. Lett.}  
\href{https://doi.org/10.1103/PhysRevLett.127.197701}{{\bf127} 197701} 

\bibitem{Parker_2021}Parker D, Ledwith P, Khalaf E, Soejima T, Hauschild J, Xie Y, Pierce A, Zaletel M P, Yacoby A and Vishwanath A 2021 arXiv:\href{https://doi.org/10.48550/arXiv.2112.13837}{2112.13837}

\bibitem{Cao2018_2}Cao Y, Fatemi V, Demir A, Fang S, Tomarken S L, Luo J Y, Sanchez-Yamagishi J D, Watanabe K, Taniguchi T, Kaxiras E, Ashoori R C and Jarillo-Herrero P 2018 {\it Nature} \href{https://doi.org/10.1038/nature26154}{{\bf556} 80–84}

\bibitem{Sharpe2019}Sharpe A L, Fox E J, Barnard A W, Finney J, Watanabe K, Taniguchi T, Kastner M A and Goldhaber-Gordon D 2019 {\it Science} 
\href{https://doi.org/10.1126/science.aaw3780}{{\bf365} 605–8}

\bibitem{Nuckolls2020}Nuckolls K P, Oh M, Wong D, Lian B, Watanabe K, Taniguchi T, Bernevig B A and Yazdani A 2020 {\it Nature} 
\href{https://doi.org/10.1038/s41586-020-3028-8}{{\bf588} 610–5}

\bibitem{Serlin2020}Serlin M, Tschirhart C L, Polshyn H, Zhang Y, Zhu J, Watanabe K, Taniguchi T, Balents L and Young A F 2020 {\it Science} 
\href{https://doi.org/10.1126/science.aay5533}{{\bf367} 900–3} 

\bibitem{Choi2020}Choi Y, Kim H, Peng Y, Thomson A, Lewandowski C, Polski R, Zhang Y, Arora H S, Watanabe K, Taniguchi T, Alicea J and Nadj-Perge S 2020 arXiv:\href{https://doi.org/10.48550/arXiv.2008.11746}{2008.11746}

\bibitem{Sharpe2021}Sharpe A L, Fox E J, Barnard A W, Finney J, Watanabe K, Taniguchi T, Kastner M A and Goldhaber-Gordon D 2021 {\it Nano Lett.} 
\href{https://doi.org/10.1021/acs.nanolett.1c00696}{{\bf21} 4299–304}

\bibitem{Andrei_2021}Wu S, Zhang Z, Watanabe K, Taniguchi T and Andrei E Y 2021 {\it Nat. Mater.} 
\href{https://doi.org/10.1038/s41563-020-00911-2}{{\bf20} 488–94}

\bibitem{Jarillo-Herrero_2021}Park J M, Cao Y, Watanabe K, Taniguchi T and Jarillo-Herrero P 2021 {\it Nature} 
\href{https://doi.org/10.1038/s41586-021-03366-w}{{\bf592} 43–8}

\bibitem{Efetov_2021}Das I, Lu X, Herzog-Arbeitman J, Song Z-D, Watanabe K, Taniguchi T, Bernevig B A and Efetov D K 2021 {\it Nat. Phys.}
\href{https://doi.org/10.1038/s41567-021-01186-3}{{\bf17} 710–4}

\bibitem{Calugaru_2022}Calugaru D, Regnault N, Oh M, Nuckolls K P, Wong D, Lee R L, Yazdani A, Vafek O and Bernevig B A 2022 {\it Phys. Rev. Lett.} 
\href{https://doi.org/10.1103/PhysRevLett.129.117602}{{\bf129} 117602}


\bibitem{Fabrizio_2022}Blason A and Fabrizio M 2022 {\it Phys. Rev.} B 
\href{https://doi.org/10.1103/PhysRevB.106.235112}{{\bf106} 235112}



\bibitem{Kwan_2021}Kwan Y H, Wagner G, Soejima T, Zaletel M P, Simon S H, Parameswaran S A and Bultinck N 2021 {\it Phys. Rev.} X 
\href{https://doi.org/10.1103/PhysRevX.11.041063}{{\bf11} 041063}

\bibitem{Kang_2018}Kang J and Vafek O 2018 {\it Phys. Rev.} X 
\href{https://doi.org/10.1103/PhysRevX.8.031088}{{\bf8} 031088}

\bibitem{Kang_2019}Kang J and Vafek O 2019 {\it Phys. Rev. Lett.} 
\href{https://doi.org/10.1103/PhysRevLett.122.246401}{{\bf122} 246401}

\bibitem{YHZhang_2019}Zhang Y-H, Mao D and Senthil T 2019 {\it Phys. Rev. Res.} 
\href{https://doi.org/10.1103/PhysRevResearch.1.033126}{{\bf1} 033126}

\bibitem{Senthil_2020}Repellin C, Dong Z, Zhang Y-H and Senthil T 2020 {\it Phys. Rev. Lett.} 
\href{https://doi.org/10.1103/PhysRevLett.124.187601}{{\bf124} 187601}

\bibitem{Dai_2021}Liu J and Dai X 2021 {\it Phys. Rev.} B 
\href{https://doi.org/10.1103/PhysRevB.103.035427}{{\bf103} 035427}

\bibitem{SongTBG3_2021}Bernevig B A, Song Z-D, Regnault N and Lian B 2021 {\it Phys. Rev.} B 
\href{https://doi.org/10.1103/PhysRevB.103.205413}{{\bf103} 205413}

\bibitem{SongTBG4_2021}Lian B, Song Z-D, Regnault N, Efetov D K, Yazdani A and Bernevig B A 2021 {\it Phys. Rev.} B 
\href{https://doi.org/10.1103/PhysRevB.103.205414}{{\bf103} 205414}

\bibitem{SongTBG5_2021}Bernevig B A, Lian B, Cowsik A, Xie F, Regnault N and Song Z-D 2021 {\it Phys. Rev.} B 
\href{https://doi.org/10.1103/PhysRevB.103.205415}{{\bf103} 205415}

\bibitem{Wu_2018}Wu F, MacDonald A H and Martin I 2018 {\it Phys. Rev. Lett.} 
\href{https://doi.org/10.1103/PhysRevLett.121.257001}{{\bf121} 257001}

\bibitem{Lian2019}Lian B, Wang Z and Bernevig B A 2019 {\it Phys. Rev. Lett.} 
\href{https://doi.org/10.1103/PhysRevLett.122.257002}{{\bf122} 257002}

\bibitem{Fabrizio_2019}Angeli M, Tosatti E and Fabrizio M 2019 {\it Phys. Rev.} X 
\href{https://doi.org/10.1103/PhysRevX.9.041010}{{\bf9} 041010}

\bibitem{Guinea_2021}Cea T and Guinea F 2021 {\it Proceedings of the National Academy of Sciences} 
\href{https://doi.org/10.1073/pnas.2107874118}{{\bf118} e2107874118 }

\bibitem{Chen_2023}Chen C, Nuckolls K P, Ding S, Miao W, Wong D, Oh M, Lee R L, He S, Peng C, Pei D, Li Y, Zhang S, Liu J, Liu Z, Jozwiak C, Bostwick A, Rotenberg E, Li C, Han X, Pan D, Dai X, Liu C, Bernevig B A, Wang Y, Yazdani A and Chen Y 2023 arXiv:\href{https://doi.org/10.48550/arXiv.2303.14903}{2303.14903}

\bibitem{LiuCX_2023}Liu C-X, Chen Y, Yazdani A and Bernevig B A 2024 {\it Phys. Rev.} B
\href{https://doi.org/10.1103/PhysRevB.110.045133 }{{\bf110} 045133 }

\bibitem{Dai_2024}Shi H, Miao W and Dai X 2024 arXiv:\href{https://doi.org/10.48550/arXiv.2402.11824}{2402.11824}

\bibitem{Song_Lian_2024}Wang Y-J, Zhou G-D, Lian B and Song Z-D 2024 arXiv:\href{https://doi.org/10.48550/arXiv.2407.11116 }{2407.11116 }

\bibitem{Song_Lian_2_2024}Wang Y-J, Zhou G-D, Peng S-Y, Lian B and Song Z-D 2024 arXiv:\href{https://doi.org/10.48550/arXiv.2402.00869}{2402.00869 }

\bibitem{Kwan_2023}Kwan Y H, Wagner G, Bultinck N, Simon S H, Berg E and Parameswaran S A 2024 {\it Phys. Rev.} B
\href{https://doi.org/10.1103/PhysRevB.110.085160}{{\bf110} 085160} 

\bibitem{Ochoa_2019}Ochoa H 2019 {\it Phys. Rev.} B 
\href{https://doi.org/10.1103/PhysRevB.100.155426}{{\bf100} 155426}

\bibitem{Ishizuka_2021}Ishizuka H, Fahimniya A, Guinea F and Levitov L 2021 {\it Nano Lett.} 
\href{https://doi.org/10.1021/acs.nanolett.1c00565}{{\bf21} 7465–71}

\bibitem{Wu_2023}Davis S M, Wu F and Sarma S D 2023 {\it Phys. Rev.} B 
\href{https://doi.org/10.1103/PhysRevB.107.235155}{{\bf107} 235155}

\bibitem{Kohn1959}Kohn W 1959 {\it Phys. Rev. Lett.} 
\href{https://doi.org/10.1103/PhysRevLett.2.393}{{\bf 2} 393–4}

\bibitem{Graphene2008}Tse W-K, Hu B Y-K and Sarma S D 2008 {\it Phys. Rev. Lett.} 
\href{https://doi.org/10.1103/PhysRevLett.101.066401}{{\bf101} 066401}

\bibitem{Nguyen2020}Nguyen T, Han F, Andrejevic N, Pablo-Pedro R, Apte A, Tsurimaki Y, Ding Z, Zhang K, Alatas A, Alp E E, Chi S, Fernandez-Baca J, Matsuda M, Tennant D A, Zhao Y, Xu Z, Lynn J W, Huang S and Li M 2020 {\it Phys. Rev. Lett.} 
\href{https://doi.org/10.1103/PhysRevLett.124.236401}{{\bf124} 236401}

\bibitem{Ando2006}Ando T 2006 {\it J. Phys. Soc. Jpn.} 
\href{https://doi.org/10.1143/JPSJ.75.12470}{{\bf 75} 124701}

\bibitem{Pimenta2009}Mafra D L, Malard L M, Doorn S K, Htoon H, Nilsson J, Castro Neto A H and Pimenta M A 2009 {\it Phys. Rev.} B 
\href{https://doi.org/10.1103/PhysRevB.80.241414}{{\bf80} 241414}

\bibitem{Jorio2022}Gadelha A C, Nadas R, Barbosa T C, Watanabe K, Taniguchi T, Campos L C, Raschke M B and Jorio A 2022 {\it 2D Mater.} 
\href{https://dx.doi.org/10.1088/2053-1583/ac8e7f}{{\bf9} 045028}

\bibitem{Ishikawa2004}Baron A Q R, Uchiyama H, Tanaka Y, Tsutsui S, Ishikawa D, Lee S, Heid R, Bohnen K-P, Tajima S and Ishikawa T 2004 {\it Phys. Rev. Lett.} 
\href{https://doi.org/10.1103/PhysRevLett.92.197004}{{\bf92} 197004}

\bibitem{Sato1991}Pouget J P, Hennion B, Escribe-Filippini C and Sato M 1991 {\it Phys. Rev.} B 
\href{https://doi.org/10.1103/PhysRevB.43.8421}{{\bf43} 8421–30 }

\bibitem{Castro2007}Lopes dos Santos J M B, Peres N M R and Castro Neto A H 2007 {\it Phys. Rev. Lett.} 
\href{https://doi.org/10.1103/PhysRevLett.99.256802}{{\bf 99} 256802}

\bibitem{MacDonald2011}Bistritzer R and MacDonald A H 2011 {\it Proc. Natl. Acad. Sci. U.S.A.} 
\href{https://doi.org/10.1073/pnas.1108174108}{{\bf 108} 12233–7}

\bibitem{Vishwanath_2019}Tarnopolsky G, Kruchkov A J and Vishwanath A 2019 {\it Phys. Rev. Lett.} 
\href{https://doi.org/10.1103/PhysRevLett.122.106405}{{\bf122} 106405}

\bibitem{SongTBG1}Bernevig B A, Song Z-D, Regnault N and Lian B 2021 {\it Phys. Rev.} B 
\href{https://doi.org/10.1103/PhysRevB.103.205411}{{\bf103} 205411}

\bibitem{Song_2022}Song Z-D and Bernevig B A 2022 {\it Phys. Rev. Lett.} 
\href{https://doi.org/10.1103/PhysRevLett.129.047601}{{\bf129} 047601}

\bibitem{Dai_2022}Shi H and Dai X 2022 {\it Phys. Rev.} B 
\href{https://doi.org/10.1103/PhysRevB.106.245129}{{\bf106} 245129}

\bibitem{Castro2011}de Gail R, Goerbig M O, Guinea F, Montambaux G and Castro Neto A H 2011 {\it Phys. Rev.} B 
\href{https://doi.org/10.1103/PhysRevB.84.045436}{{\bf84} 045436}

\bibitem{He2013}He W-Y, Chu Z-D and He L 2013 {\it Phys. Rev. Lett.} 
\href{https://doi.org/10.1103/PhysRevLett.111.066803}{{\bf111} 066803}

\bibitem{Po2019}Po H C, Zou L, Senthil T and Vishwanath A 2019 {\it Phys. Rev.} B 
\href{https://doi.org/10.1103/PhysRevB.99.195455}{{\bf99} 195455}

\bibitem{Koshino2020}Koshino M and Nam N N T 2020 {\it Phys. Rev.} B 
\href{https://doi.org/10.1103/PhysRevB.101.195425}{{\bf101} 195425}

\bibitem{LiuJP2023}Xie B and Liu J 2023 {\it Phys. Rev.} B 
\href{https://doi.org/10.1103/PhysRevB.108.094115}{{\bf108} 094115}

\bibitem{Mariani2008}Mariani E and von Oppen F 2008 {\it Phys. Rev. Lett.} 
\href{https://doi.org/10.1103/PhysRevLett.100.076801}{{\bf100} 076801}

\bibitem{Ando2002}Suzuura H and Ando T 2002 {\it Phys. Rev.} B 
\href{https://doi.org/10.1103/PhysRevB.65.235412}{{\bf65} 235412}

\bibitem{Pereira2009}Pereira V M and Castro Neto A H 2009 {\it Phys. Rev. Lett.} 
\href{https://doi.org/10.1103/PhysRevLett.103.046801}{{\bf103} 046801}

\bibitem{Guinea2010}Guinea F, Katsnelson M I and Geim A K 2010 {\it Nature Phys} 
\href{https://doi.org/10.1038/nphys1420}{{\bf6} 30–3}

\bibitem{Mariani2010}Mariani E and Von Oppen F 2010 {\it Phys. Rev.} B 
\href{https://doi.org/10.1103/PhysRevLett.101.096802}{{\bf82} 195403}

\bibitem{Ochoa2012}Ochoa H, Castro E V, Katsnelson M I and Guinea F 2012 {\it Physica E: Low-dimensional Systems and Nanostructures} 
\href{https://doi.org/10.1016/j.physe.2011.03.017}{{\bf44} 963–6}

\end{thebibliography}
\end{document}